\documentclass{style/nseJournal}

\usepackage{graphicx} 
\usepackage{booktabs} 
\usepackage{microtype} 
\usepackage{physics}
\usepackage{amsmath}
\usepackage{siunitx}
\usepackage{bm}
\usepackage[version=4]{mhchem}
\usepackage{isotope}

\newcommand{\keff}{k_{\text{eff}}}
\newcommand{\ktot}{k_{\text{tot}}}
\newcommand{\pcm}{\text{pcm}}
\newcommand{\Pmu}{P_{g' \rightarrow g}(\mu)}
\newcommand{\Cmu}{C_{g' \rightarrow g}(\mu)}
\newcommand{\dir}{\bm{\hat{\Omega}}}
\newcommand{\pos}{\bm{r}}
\newcommand{\Et}{\Sigma_t}
\newcommand{\Es}{\Sigma_s}
\newcommand{\Ef}{\Sigma_f}

\newcommand{\varphip}{\varphi_{+}}
\newcommand{\varphin}{\varphi_{-}}
\newcommand{\varphitot}{\varphi_{\text{tot}}}
\newcommand{\Esp}{\Sigma_{s+}}
\newcommand{\Esn}{\Sigma_{s-}}
\newcommand{\scatt}{\bm{\hat{S}}}
\newcommand{\fiss}{\bm{\hat{F}}}
\newcommand{\cm}{\,\text{cm}}

\begin{document}

\title{Negative Weights and Weight Cancellation to Treat Anisotropic Scattering in Multigroup Monte Carlo Simulations} 

\addAuthor{\correspondingAuthor{Parth Singh}}{a}
\addAuthor{Hunter Belanger}{a}
\correspondingEmail{singhp10@rpi.edu}

\addAffiliation{a}{Department of Mechanical, Aerospace, and Nuclear Engineering, \\
                   Rensselaer Polytechnic Institute, 110 8th St., Troy, NY, USA}

\addKeyword{Anisotropic Scattering}
\addKeyword{Negative weights}
\addKeyword{Weight Cancellation}
\addKeyword{Monte Carlo}
\addKeyword{Multigroup}

\titlePage

\begin{abstract}
The Monte Carlo method is typically considered the gold standard for simulating reactor physics problems, as it does not require discretization of the phase space. This is not necessarily true though when simulating multigroup problems, as it has traditionally been a challenge to model anisotropic scattering in such simulations. Multigroup data used in reactor simulations generally uses low order Legendre expansion for scattering distributions, often stopping at the third Legendre moment. With so few terms, the angular distribution can easily have negative regions for highly anisotropic energy transfers, which makes it impossible to use standard Monte Carlo methods to sample a scattering angle. Multigroup Monte Carlo codes therefore often resort to only using isotropic scattering with the transport correction (which is not always possible), or approximating the angular distribution with discrete angles. Neither case is ideal, and makes it impossible to accurately model such problems or verify deterministic codes that can and do make use of the low order Legendre expansions without issue. This is addressed in the present study by using importance sampling in conjunction with negative particle weights to sample scattering angles from negative scattering distributions. It is demonstrated that such an approach necessitates the use of weight cancellation methods in order to be stable and converge to a solution. The technique is tested on two simple analytic benchmark problems, and then further demonstrated by modeling a small zero power research reactor, comparing results against a deterministic solver which can treat anisotropic scattering. Comparison of the simulation results indicates that importance sampling for anisotropic scattering with weight cancellation can be used to obtain reference Monte Carlo results for multigroup problems despite negative scattering distributions.
\end{abstract}

\section{Introduction}
In the current era of powerful and affordable personal computers, the Monte Carlo method for solving the neutron transport equation has practically become synonymous with continuous energy, where the complex energetic behavior of isotopic cross sections can be modeled as a continuous function, avoiding discretization. This is, of course, one of the main advantages of the method, as it permits one to forego the intricate and nuanced procedure of generating appropriate multigroup cross sections for a given material and geometric configuration, which is a problem of primary interest when solving the neutron transport equation with deterministic methods \cite{knott2010}. Not so long ago, however, it was somewhat common place to have multigroup Monte Carlo transport codes. You have the benefits of a continuous representation of position and direction, and discretizing along the energy variable afforded faster run times and lower memory consumption on historic computers, when such concerns were at the forefront of a developer's mind. Examples of such codes include MORET 5, TRIPOLI-3, KENO-V, just to name a small handful \cite{MORET5, T3, kenov}.

Very recent works, however, have highlighted new applications of multigroup Monte Carlo simulations. The first is the advent of GPU computing architectures, which are quickly replacing traditional CPU based HPC platforms at national laboratories. GPUs use a vectorized computing methodology based on the paradigm of single-instruction multiple data (SIMD). These systems are therefore most efficient when all threads are performing the same set of machine instructions, but on different pieces of data. Unfortunately, the polymorphic nature of continuous energy nuclear data makes such a SIMD approach to sampling neutron reactions difficult, as each distribution in the ENDF format has a particular algorithm for sampling the outgoing angle and energy of the scattered particle \cite{ENDF_manual}. The multigroup formalism does not have this problem as a scattering matrix can be sampled in the same manner, regardless of the reaction being considered. This makes multigroup Monte Carlo trivially vectorizable, and an excellent candidate for implementation on GPU architectures. Examples of this are the Profugus multigroup Monte Carlo code developed at Oak Ridge National Laboratories for GPUs \cite{Hamilton2018}, and the MGMC code developed at Los Alamos National Laboratory \cite{BurkeMGMC}. Another recent use of multigroup Monte Carlo is for the acceleration of the fission source convergence \cite{Raffuzzi2023}. Since multigroup simulations run much faster than continuous energy simulations, one could quickly allow the fission source to converge to the fundamental eigenstate before moving back to continuous energy for the portion of the simulation where tally statistics are generated. Raffuzi et al.\ demonstrated this method to be very effective for LWR systems, achieving speed-ups between a factor of 2.5 and 5 \cite{Raffuzzi2023, Raffuzzi2023PhD}.

One of the difficulties of using multigroup Monte Carlo as outlined by Raffuzzi \cite{Raffuzzi2023PhD} and Burke \cite{BurkeMGMC} is the treatment of anisotropic scattering. Traditionally, angular distributions are represented with a truncated Legendre expansion, often only going up to the 3rd order. Such a truncation often results in probability distributions for the scattering angle which are negative, and therefore nonphysical. This has traditionally made it impossible to sample a scattering angle in a multigroup Monte Carlo code using the ``true'' truncated distribution. Most multigroup Monte Carlo codes circumvent this problem by using a discrete angle representation which preserves the Legendre moments of the scattering distribution \cite{MORET5,kenov,BurkeMGMC,Raffuzzi2023PhD}. Despite preserving the Legendre moments, the discrete angle representation is not equivalent to the continuous truncated Legendre distribution, and can give very different results (as will be demonstrated later in this paper). This limitation of the anisotropic scattering representation is therefore a large barrier for implementing multigroup Monte Carlo codes which can reproduce continuous energy reference solutions. This also makes it impossible to adequately verify deterministic codes that use an explicit anisotropic scattering representation, as it is currently impossible to obtain reference Monte Carlo solutions. An alternative approximation could be to use the transport correction and then employ isotropic scattering \cite{Stammler1983}. This can be problematic in its own right, however, as one must choose between the inflow and outflow approximations for the transport correction \cite{Imapact_of_inflow_tr}. Additionally, the transport correction can result in negative diagonal elements of the scattering matrix, or even negative transport cross sections \cite{Stammler1983}. This then requires other specialized methods in order to sample a flight distance and energy transfer in a Monte Carlo code when using the transport correction approximation \cite{Liu2023}.

The objective of this work is therefore to present an algorithm based on importance sampling which permits one to sample scattering angles from a truncated Legendre distribution with negative regions in an exact manner, as will be shown in Sec.~\ref{sec:importance_sampling}. This technique will require that particles be allowed to carry a negative statistical weight, which has long been known to prevent the convergence of criticality simulations. In Sections~\ref{sec:stability} and \ref{sec:cancellation}, it will be demonstrated that a ``weight cancellation'' procedure must be applied to the fission source, where positive and negative weighted particles annihilate with one another \cite{exact_weight_cancellation}, in order for the simulation to be stable and converge to the solution. Our proposed methodology is tested on two simple test problems with analytically known eigenvalues in Sec.~\ref{sec:test_problems}. Sec.~\ref{sec:RCF} applies the method to a zero power research reactor simulation, and compares the results against different transport corrections approximations. Finally, concluding remarks and a brief summary of the results are provided in Sec.~\ref{sec:conclusions}.

\section{Methods and Theory}\label{sec:methods_theory}
\begin{figure}
    \centering
    \includegraphics[width=0.55\linewidth]{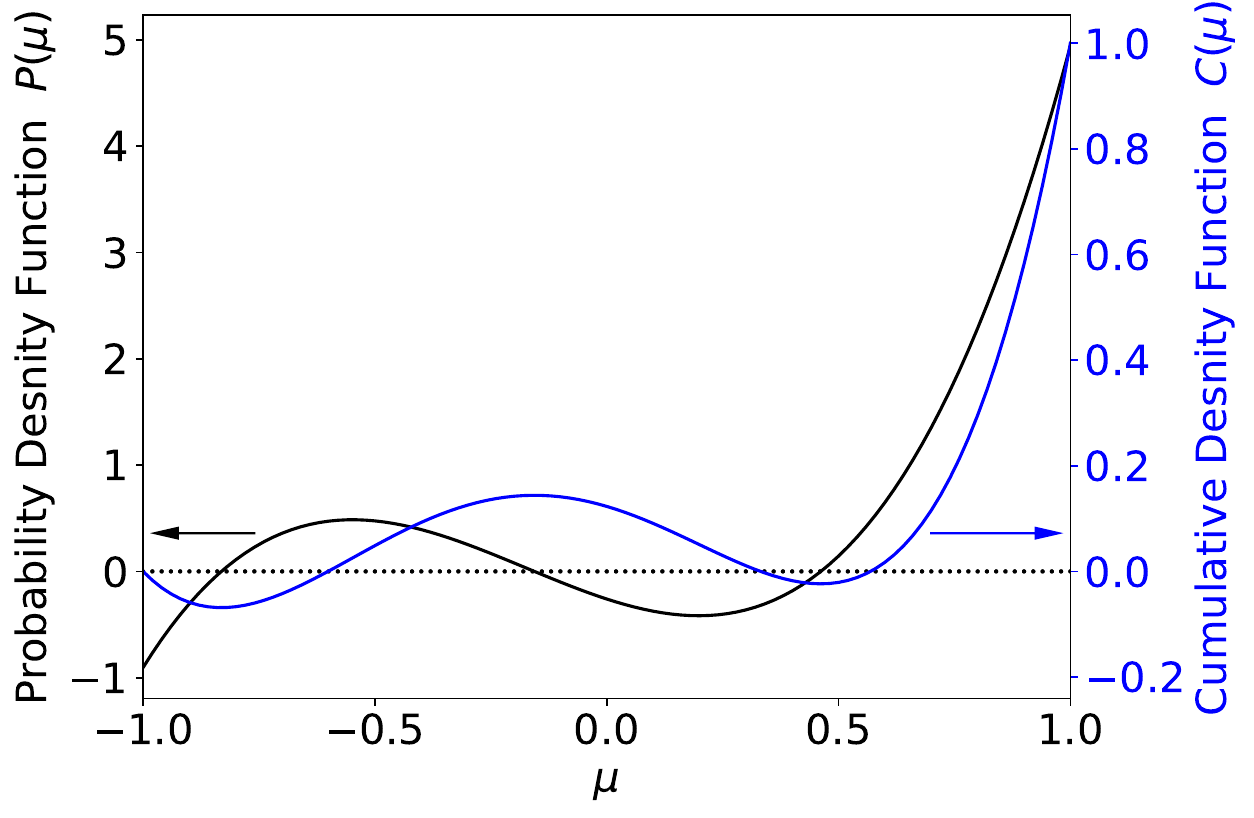}
    \caption{Scattering Distribution and its CDF for in-group scattering within the energy interval $\SI{9.999987}{\mega\eV} - \SI{19.6403}{\mega\eV}$ in water with $L=3$ for RPI's Research Critical Facility (RCF).}
    \label{fig:fig1-RCF-H2O-g1-PDF-CDF}
\end{figure}

Scattering angular distributions of neutrons for any group-to-group energy transfer are often represented as a finite series expansion of Legendre polynomials for the cosine of the scattering angle, $\mu$. The sampling of $\mu$ can be performed from its group-to-group transfer scattering distribution $\Pmu$, written as
\begin{equation}{\label{eq:scattering-series-expansion}}
    \Pmu = \frac{1}{\Sigma_{s,0, g' \rightarrow g}} \sum_{l=0}^{L}
    \frac{2l+1}{2}
    p_{l}(\mu)
    \Sigma_{s,l, g' \rightarrow g} 
\end{equation}
where, $\mu$ is cosine of the scattering angle, $\Sigma_{s,l, g' \rightarrow g}$ is the $l$th Legendre moment of the scattering cross section from group $g'$ to group $g$, and $p_l(\mu)$ is the Legendre Polynomial of the $l$th order. As long as the $L$ is sufficiently large, the distribution $\Pmu$ obeys the fundamental laws of probability theory, implying $\Pmu\geq 0 \text{ } \forall\ \mu \in [-1, 1]$ and $C_{g'\rightarrow g}(1) = 1$, where $C_{g'\rightarrow g}(\mu)$ is the cumulative density function (CDF)
\begin{equation}
    C_{g'\rightarrow g}(\mu) = \int_{-1}^{\mu}P_{g' \rightarrow g}(\mu')\mathrm{d}\mu'.
\end{equation}
With a random number $\xi\sim \mathcal{U}(0,1)$, one can directly sample a value of $\mu$ using 
\begin{equation}
    \mu = C^{-1}_{g'\rightarrow g}(\xi),
\end{equation}
where $C^{-1}_{g'\rightarrow g}$ is the inverse function of $C_{g'\rightarrow g}$. Since $\Cmu$ is monotonically increasing, there is a unique value of $\mu$ for a particular value of $\xi$. If, however, the Legendre series expansion is truncated at too small value of $L$, the approximated angular distribution may have negative regions over the interval of $\mu \in [-1, 1]$. Fig.\ref{fig:fig1-RCF-H2O-g1-PDF-CDF} shows the angular distribution and the CDF of the in-group scattering for the energy interval $\SI{9.999987}{\mega\eV} - \SI{19.6403}{\mega\eV}$ in water with $L=3$ scattering moments for the RPI Research Critical Facility (RCF). This energy transfer is highly anisotropic and the scattering distribution has two negative regions over the interval of $[-1, 1]$. In such cases, the monotonicity of $C_{g'\rightarrow g}$ is lost, resulting in multiple values of $\mu$ for particular values of $\xi$. As a consequence, the sampling of $\mu$ is no longer well defined because $\Pmu$ is no longer statistically valid.  
One could certainly eliminate these negative regions simply by increasing the order of scattering moments, $L$, until $P_{g'\rightarrow g}(\mu) \geq 0 \text{ } \forall \mu \in [-1, 1]$. However, increasing the order of the series expansion is not always desirable as the memory requirement to store the scattering matrices is directly proportional to $L$. Often, deterministic codes do not exceed $L=3$ for most of light water calculations \cite{knott2010}. Having negative regions in the scattering distributions does not affect deterministic codes, but makes it impossible to run the equivalent problem with a Monte Carlo code to obtain a reference solution. 

\subsection{Importance Sampling}\label{sec:importance_sampling}
The present work focuses on the sampling of scattering angles from a distribution which has negative regions, and employs the well known method of importance sampling. The true distribution that we would like to sample shall be denoted as $P_{g'\rightarrow g}(\mu)$, and has negative regions for $\mu\in[-1,1]$. As outlined above, these negative regions make it impossible to directly sample a scattering angle through a method such as inversion. Instead, we shall assume that there is another angular distribution, $P'_{g' \rightarrow g}(\mu)$, that obeys the fundamental laws of probability theory (i.e.\ is normalized and positive $\forall\mu\in [-1,1]$), from which a scattering angle can be readily sampled using standard methods. In a Monte Carlo simulation, particles often carry a statistical weight, $w$, that is used as a multiplying factor when contributing to the estimation of observables such as the flux or reaction rates \cite{lux_koblinger_MC_book_1991}. It is almost a universal rule of thumb for Monte Carlo developers that this statistical weight be positive, and it is generally assumed that $w\in(0,1]$, though values greater than unity are often allowed with variance reduction methods such as weight windows and FW-CADIS \cite{Wagner2017}. The idea of importance sampling is that one may sample a scattering angle from $P'_{g'\rightarrow g}(\mu)$ \emph{instead of} sampling from the true distribution $P_{g'\rightarrow g}(\mu)$, so long as the particle's statistical weight is adjusted in a manner that keeps the simulation unbiased (the expectation values of observable quantities do not change).

The general sampling strategy in a multigroup formalism then takes the following form. When a particle undergoes a scattering event with incident energy group $g'$, the energy transition $g'\to g$ is first sampled in the standard manner using the $P0$ scattering matrix\footnote{In this notation, $PN$ corresponds to a Legendre expansion with $L=N$ Legendre moments.}. Once this energy transition is known, if the true scattering distribution has negative regions, the scattering angle is sampled from the alternative distribution $P'_{g'\rightarrow g}(\mu)$ using standard inversion techniques. Subsequently, to keep the game unbiased, the particle's weight is multiplied by the likelihood ratio, which is the ratio of the true scattering distribution to the alternative distribution, $P_{g'\rightarrow g}(\mu) / P'_{g'\rightarrow g}(\mu)$, evaluated at the sampled value of $\mu$. It is important to note that while $P'_{g'\rightarrow g}(\mu)$ is always positive, $P_{g'\rightarrow g}(\mu)$ is not, so particles are allowed to have positive or negative statistical weights, and can change between the two signs upon multiple scattering events. This method can of course be used with other weighted variance reduction techniques such as implicit capture and forced fission \cite{lux_koblinger_MC_book_1991}.

Choosing an alternative distribution, $P'_{g'\rightarrow g}(\mu)$, is one of the key aspects of importance sampling. In this work, the suggestion of Brockmann is used \cite{brockmann1981}, where
\begin{equation}{\label{eq:modified_PDF}}
    P'_{g'\rightarrow g}(\mu) = \frac{\abs{P_{g'\rightarrow g}(\mu)}}{\displaystyle\int_{-1}^{1}\abs{P_{g'\rightarrow g}(\mu')}\mathrm{d}{\mu'}}.
\end{equation}
This choice will lead to a particle weight multiplier of
\begin{equation}{\label{eq:weight_modifier_ratio}}
    \frac{P_{g'\rightarrow g}(\mu)}{P'_{g'\rightarrow g}(\mu)} =
    \frac{P_{g'\rightarrow g}(\mu)}{\abs{P_{g'\rightarrow g}(\mu)}}
    \int_{-1}^{1}\abs{P_{g'\rightarrow g}(\mu')}\dd\mu' =
    \mathrm{sgn}\left({P_{g'\rightarrow g}(\mu)}\right)\int_{-1}^{1}\abs{P_{g'\rightarrow g}(\mu')}\dd\mu'.
\end{equation}
The magnitude of this likelihood ratio is therefore independent of $\mu$, and the sign is taken from the true angular distribution, $P_{g'\rightarrow g}(\mu)$. The modified angular distribution given by Eq.~\eqref{eq:modified_PDF} will closely resemble the true angular distribution, so long as there are no excessively negative regions in the true distribution. A special case of the choice of the distribution is if the true angular distribution is positive, then the alternative distribution will be the same as true angular distribution with a likelihood ratio of unity.

It is important to point out that the use of importance sampling to handle negative scattering distributions in Monte Carlo simulations is not new, and is of course considered by Brockmann in his work \cite{brockmann1981}. Importance sampling is also regularly used in other aspects of Monte Carlo simulations \cite{lux_koblinger_MC_book_1991}. Despite this, to the best of the authors' knowledge, importance sampling has never been regularly used in production Monte Carlo transport codes to treat anisotropic scattering. Two well known examples of multigroup Monte Carlo codes are the KENO-V code from Oak Ridge in the United States \cite{kenov} and the MORET 5 code from the IRSN in France \cite{MORET5}. Both codes reconstruct approximated distributions which preserve the provided Legendre moments. In KENO-V, a discrete distribution is always used, while MORET 5 uses a discrete distribution for $P5$ and above, but semi-continuous distributions for $P1$ (Coveyou's law \cite{Coveyou1965}) and $P3$ \cite{MORET5}. While the authors have not been able to find published literature addressing this, it appears that the reason importance sampling has not been historically used to treat anisotropic scattering in Monte Carlo codes is because it was observed that introducing negative weights made the simulations exceptionally unstable, they would never converge, and/or would run out of memory \cite{BrownBelangerDefense}. This problem is somewhat alluded to by Brockmann, though not necessarily with the emphasis that this point deserves \cite{brockmann1981}. Recent works have examined the stability of power iteration simulations that employ negative weights in other contexts \cite{exact_weight_cancellation}. Using a similar approach, we analyze the stability of Monte Carlo simulations that use importance sampling to treat negative scattering distributions in the subsequent section.

\subsection{Power Iteration Stability In The Presence of Negative Weights}\label{sec:stability}

Belanger et al.\ have previously demonstrated the potential benefits of using negative weights in Monte Carlo particle simulations, particularly when modeling spatially continuous material properties \cite{Belanger2020}. In that work, negative weighted delta tracking was used to sample flight distances of particles in fixed-source simulations \cite{Legrady2017, Carter1972}. However, subsequent studies demonstrated that using negative weights in power iteration simulations prevents the simulation from converging to the fundamental eigenstate \cite{exact_weight_cancellation}. In their work, the number of particles in the simulation kept increasing, resulting in increases in the time to process each fission generations as well as increases in the memory required to store the particles \cite{exact_weight_cancellation}. Many population control mechanisms, such as combing and splitting, were tested by Belanger et al., but none were able to control the particle population, and simulations were eventually killed by the operating system due to a lack of system resources\cite{exact_weight_cancellation}. Their work provided a detailed derivation as to why power iteration simulations will fail when using the negative weighted delta tracking algorithm. In this section, we provide a similar derivation which demonstrates why power iteration with negative weights introduced by importance sampling for angular distributions will similarly fail to converge. The derivation presented here is performed in continuous energy, however, it is trivial to modify it to a multigroup formalism.

To begin, we start with the k-eigenvalue Boltzmann transport equation which we write as
\begin{equation}\label{eq:boltzmann_op}
    \dir\cdot\grad \varphi(\pos,E,\dir) + \Et(\pos,E) \varphi(\pos,E,\dir) 
    = \frac{1}{k} \fiss\varphi(\pos, E, \dir) + \scatt\varphi(\pos, E, \dir).
\end{equation}
Here, $\fiss$ is the fission operator, defined as
\begin{equation}{\label{eq:fission_operator}}
    \fiss\varphi(\pos, E, \dir) = \frac{\chi(\pos, E)}{4\pi} \iint \nu\Ef(\pos, E') \varphi(\pos,E',\dir') {\dd \dir'}{\dd E'},
\end{equation}
and $\scatt$ is the scattering operator, defined as
\begin{equation}{\label{eq:scattering_operator}}
    \scatt\varphi(\pos, E, \dir) = \iint \Es(\pos, E'\rightarrow E, \dir' \rightarrow \dir)\varphi(\pos,E',\dir'){\dd\dir'}{\dd E'}.
\end{equation}
Throughout this work, $\varphi$ is the angular flux, $\Et$ is the total cross section, $k$ is the eigenvalue, $\chi$ is the fission energy spectrum, $\nu$ is the average number of the neutrons released per fission, $\Ef$ is the fission cross section, and $\Es$ is the scattering cross section.

In the present work, negative weights are introduced through importance sampling on the scattering distribution. If a scattering angle is sampled from the negative region of a scattering distribution, the particle's statistical weight will change sign once multiplied by the likelihood ratio given in Eq.~\eqref{eq:weight_modifier_ratio}. Therefore, phase space regions with a negative scattering distribution couple the positive and negative particle populations. To mathematically represent this coupling, the scattering kernel can be split into the two parts based on the sign of the angular distribution:
\begin{equation}{\label{eq:Es_pos_neg_eq}}
        \Es(\pos, E'\rightarrow E, \dir' \rightarrow \dir) 
        = \Esp(\pos, E'\rightarrow E, \dir' \rightarrow \dir) 
        - \Esn(\pos, E'\rightarrow E, \dir' \rightarrow \dir),
\end{equation}
where $\Esp$ and $\Esn$ are the scattering cross sections for the positive regions and negative regions, respectively. These terms are defined as   
\begin{equation}{\label{eq:Es_positive}}
    \Esp(\pos, E'\rightarrow E, \dir' \rightarrow \dir) \equiv \Es(\pos, E'\rightarrow E, \dir' \rightarrow \dir)\mathcal{H}\left (\Es(\pos, E'\rightarrow E, \dir' \rightarrow \dir)\right ) 
\end{equation}
and
\begin{equation}{\label{eq:Es_negative}}
    \Esn(\pos, E'\rightarrow E, \dir' \rightarrow \dir) \equiv \abs{\Es(\pos, E'\rightarrow E, \dir' \rightarrow \dir)}\mathcal{H}\left(-\Es(\pos, E'\rightarrow E, \dir' \rightarrow \dir)\right),
\end{equation}
where $\mathcal{H}(x)$ is the Heaviside step function:
\begin{equation}{\label{eq:heaviside_function}}
    \mathcal{H}(x) = 
    \begin{cases}
        1 & \text{if } x \geq 0 \\
        0 & \text{if } x < 0
    \end{cases}.
\end{equation}
It is important to note that while $\Esn$ represents the negative portions of the scattering distribution, $\Esn$ itself is a strictly positive quantity and has the desired effect thanks to the negative sign in Eq.~\eqref{eq:Es_pos_neg_eq}. Based on this decomposition, the scattering operator can also be split into two terms, first $\scatt_{+}\varphi$ is the scattering from positive regions of the distribution and $\scatt_{-}\varphi$ is the scattering from negative regions. Formally, these two may be defined as
\begin{equation}{\label{eq:scattering_postive_operator}}
    \scatt_{+}\varphi(\pos, E, \dir) = \iint \Esp(\pos, E'\rightarrow E, \dir' \rightarrow \dir)\varphi(\pos,E',\dir'){\dd\dir'}{\dd E'}
\end{equation}
and
\begin{equation}{\label{eq:scattering_negative_operator}}
    \scatt_{-}\varphi(\pos, E, \dir) = \iint \Esn(\pos, E'\rightarrow E, \dir' \rightarrow \dir)\varphi(\pos,E',\dir'){\dd\dir'}{\dd E'},
\end{equation}
indicating the original scattering operator can also be written as
\begin{equation}\label{eq:scat_op_pos_neg}
    \scatt\varphi(\pos, E, \dir) = \scatt_{+}\varphi(\pos, E, \dir) - \scatt_{-}\varphi(\pos, E, \dir).
\end{equation}

Following the process used in Belanger et al., we model positive and negative particles as two separate \emph{species} which are coupled \cite{exact_weight_cancellation}. To accomplish this, we define $\varphi_{+}$ as the flux due to the positive particles and $\varphi_{-}$ as the flux due to the negative particles. Both of these quantities are strictly positive, and the physical flux of the system, $\varphi$, can be expressed in terms of $\varphi_{+}$ and $\varphi_{-}$ as
\begin{equation}{\label{eq:phi_net_eq}}
        \varphi(\pos, E, \dir) = \varphi_{+}(\pos, E, \dir) - \varphi_{-}(\pos, E, \dir).
\end{equation}
Using the scattering operators provided in Eq.~\eqref{eq:scattering_postive_operator} and~\eqref{eq:scattering_negative_operator}, we can write a set of coupled transport equations for the positive and negative particles, given in Eq.~\eqref{eq:boltzmann_postive_weight} and~\eqref{eq:boltzmann_negative_weight} respectively:
\begin{equation}{\label{eq:boltzmann_postive_weight}}
    \dir\cdot\grad \varphi_{+}(\pos,E,\dir) + \Et(\pos,E) \varphi_{+}(\pos,E,\dir) 
    = \frac{1}{k} \fiss\varphi_{+}(\pos, E, \dir) + \scatt_{+}\varphi_{+}(\pos, E, \dir) + \scatt_{-}\varphi_{-}(\pos, E, \dir)
\end{equation}
\begin{equation}{\label{eq:boltzmann_negative_weight}}
    \dir\cdot\grad \varphi_{-}(\pos,E,\dir) + \Et(\pos,E) \varphi_{-}(\pos,E,\dir) 
    = \frac{1}{k} \fiss\varphi_{-}(\pos, E, \dir) + \scatt_{+}\varphi_{-}(\pos, E, \dir) + \scatt_{-}\varphi_{+}(\pos, E, \dir).
\end{equation}
These coupled equations describe how particles move from being positive to negative, and vice versa. In our case, the coupling comes from the $\scatt_{-}$ operator, where $\scatt_{-}\varphi_{-}$ is a source of positive particles and $\scatt_{-}\varphi_{+}$ is a source of negative particles. Using the identities provided by Eq.~\eqref{eq:phi_net_eq} and~\eqref{eq:scat_op_pos_neg}, it is easy to demonstrate that the difference of Eq.~\eqref{eq:boltzmann_postive_weight} and~\eqref{eq:boltzmann_negative_weight} yields the physical Boltzmann k-eigenvalue transport equation of Eq.~\eqref{eq:boltzmann_op}.

However, considering Belanger et al.'s derivation and following suit, one could also take the sum of Eq.~\eqref{eq:boltzmann_postive_weight} and~\eqref{eq:boltzmann_negative_weight}. To this end, we define the quantity
\begin{equation}{\label{eq:phi_total_eq}}
        \varphitot(\pos, E, \dir) =  \varphip(\pos, E, \dir) + \varphin(\pos, E, \dir).
\end{equation}
Using this definition, Eq.~\eqref{eq:boltzmann_postive_weight} and~\eqref{eq:boltzmann_negative_weight} sum to
\begin{multline}{\label{eq:boltzmann_phi_total}}
    \dir\cdot\grad \varphitot(\pos,E,\dir) + \Et(\pos,E) \varphitot(\pos,E,\dir) = \\
    \frac{1}{\ktot} \fiss\varphitot(\pos, E, \dir) + \scatt_{+}\varphitot(\pos, E' \dir') + \scatt_{-}\varphitot(\pos, E' \dir'),
\end{multline}
where we have labeled the eigenvalue of this equation $\ktot$, for convenience. This equation is very similar to the Boltzmann equation given in Eq.~\eqref{eq:boltzmann_op}; the total and fission cross sections are identical. The only difference is the effective scattering cross section, which we shall denote as
\begin{equation}
    \Es{_{,\text{tot}}}(\pos, E'\rightarrow E, \dir' \rightarrow \dir) 
    = \Es{_+}(\pos, E'\rightarrow E, \dir' \rightarrow \dir)  + \Es{_-}(\pos, E'\rightarrow E, \dir' \rightarrow \dir).
\end{equation}
Because $\Es{_{,\text{tot}}}$ takes the sum of $\Esp$ and $\Esn$,
\begin{equation}\label{eq:scat_tot_larger}
    \iint \Es(\pos, E\rightarrow E', \dir\rightarrow\dir') \dd\dir'\dd E \leq
    \iint \Es{_{,\text{tot}}}(\pos, E\rightarrow E', \dir\rightarrow\dir') \dd\dir'\dd E,
\end{equation}
with equality only holding when $\Esn = 0$. As the total and fission cross sections are identical in Eqs.~\eqref{eq:boltzmann_op} and~\eqref{eq:boltzmann_phi_total}, Eq.~\eqref{eq:scat_tot_larger} implies that Eq.~\eqref{eq:boltzmann_phi_total} has a smaller sterile capture cross section than Eq.~\eqref{eq:boltzmann_op}:
\begin{equation}
    \Sigma_{c, \text{tot}}(\pos, E) = \Et(\pos, E) - \Ef(\pos, E) - \Es{_{,\text{tot}}}(\pos, E)   \leq \Et(\pos, E) - \Ef(\pos, E) - \Es(\pos, E) = \Sigma_c(\pos, E).
\end{equation}
Therefore, the multiplication factor for Eq.~\eqref{eq:boltzmann_phi_total}, $\ktot$, will be greater than the multiplication factor of the Boltzmann equation, $k$ (which is $\keff$). As a consequence, the dominant eigenvalue of the coupled system of positive and negative particles is the non-physical eigenvalue $\ktot$. Thus, so long as $\Esn > 0$, power iteration will fail to converge towards physical eigenstate corresponding to $\varphi$. 

\subsection{Weight Cancellation}\label{sec:cancellation} 

Our explanation of the failure of power iteration in the presence of negative weights due to importance sampling on the scattering distribution is completely analogous to the derivation previously performed by Belanger et al.\ to describe the failure of power iteration due to negative weights introduced by negative weighted delta tracking \cite{exact_weight_cancellation}. Their work also proposed that power iteration with negative weights could be stabilized through a weight cancellation procedure, where positive and negative weighted particles annihilate with one another. This was demonstrated in both deterministic and Monte Carlo simulations, and was shown to be effective at permitting convergence of the simulation to the physical flux, $\varphi$. Their proof for why weight cancellation permits convergence of power iteration can also be directly applied to our system where negative weights are introduced through the scattering operator. Weight cancellation effectively suppresses the eigenvalue of the non-physical system, making the eigenvalue of the physical system dominant. Belanger et al. have also shown that there is a minimum amount of weight cancellation that is required for power iteration to converge on the physical eigenstate. Therefore, if there is an insufficient amount weight cancellation, the non-physical eigenstate, $\varphitot$, will be dominant \cite{exact_weight_cancellation}.

In the present work, approximate regional weight cancellation is used as the weight cancellation algorithm to permit the convergence of power iteration to $\varphi$ \cite{solving_eigenvalue_transport_with_Neg_weight}. While exact weight cancellation procedures have been developed for multigroup simulations \cite{exact_weight_cancellation, belanger2022}, approximate weight cancellation is very simple to implement and fast to run. In the approximate regional weight cancellation method, a regular rectilinear spatial mesh is imposed over the problem domain. Between each fission generation, the fission source particles are sorted into these cancellation mesh bins. Once all particles are sorted, the average weight of the particles within a single mesh bin can be calculated. The average weight of the bin is then assigned to all particles in that bin. This method imposes the bias on the fission source, which can be reduced by refining the cancellation mesh. However, refining the cancellation mesh will cause there to be fewer particles in each bin leading to less weight cancellation \cite{solving_eigenvalue_transport_with_Neg_weight, exact_weight_cancellation}. The bias imposed by approximate cancellation is generally acceptable, because it is applied to the fission source. Neutron emission from fission is assumed to be almost perfectly isotropic, and the fission energy spectrum is only weakly dependent on the incident neutron energy. As a result, the bias is largely contained to the spatial coordinates of the fission source, and the bias on the the direction and energy coordinates of the fission source is minimal.

\section{Simple Analytic Test Problems}\label{sec:test_problems}

The methodology of importance sampling in conjunction with approximate regional weight cancellation, as presented in Sec.~\ref{sec:methods_theory}, has been implemented in the Abeille Monte Carlo code \cite{abeille_repo}. In this section, the methodology is verified against simplified criticality test problems with analytically known eigenvalues. These are problems 37 and 71 taken from Sood et al. \cite{sood2003}. Problem-37 is an infinite cylinder geometry and problem-71 is an infinite slab geometry. For both problems, the analytically known values of $\keff$ are $1$. The Monte Carlo simulations were performed using the delta-tracking transport algorithm \cite{lux_koblinger_MC_book_1991}. As mentioned in the previous section, the bias imposed by approximate weight cancellation can be reduced by refining the size of the cancellation mesh. Therefore, different regular rectilinear meshes were considered to study the convergence of $\keff$ and the flux. The Monte Carlo simulation results with importance sampling and weight cancellation were benchmarked against results from the deterministic transport library, Scarabée \cite{scarabée}. Scarabée contains several different solvers, for various types of geometries. Problem-71 is a 1D slab, so Scarabée's 1D $S_N$ solver with anisotropic scattering was used for the reference solution. As problem-37 is 2D, it was simulated using the 2D method of characteristics (MOC) solver with an explicit anisotropic scattering treatment using spherical harmonics. The estimated values of $\keff$ from Scarabée are in excellent agreement with the analytically know values, as will be seen subsequently, making Scarabée an excellent option for validating the importance sampling anisotropic scattering treatment.

\begin{figure}[hbpt]
    \centering
    \includegraphics[width=0.75\linewidth]{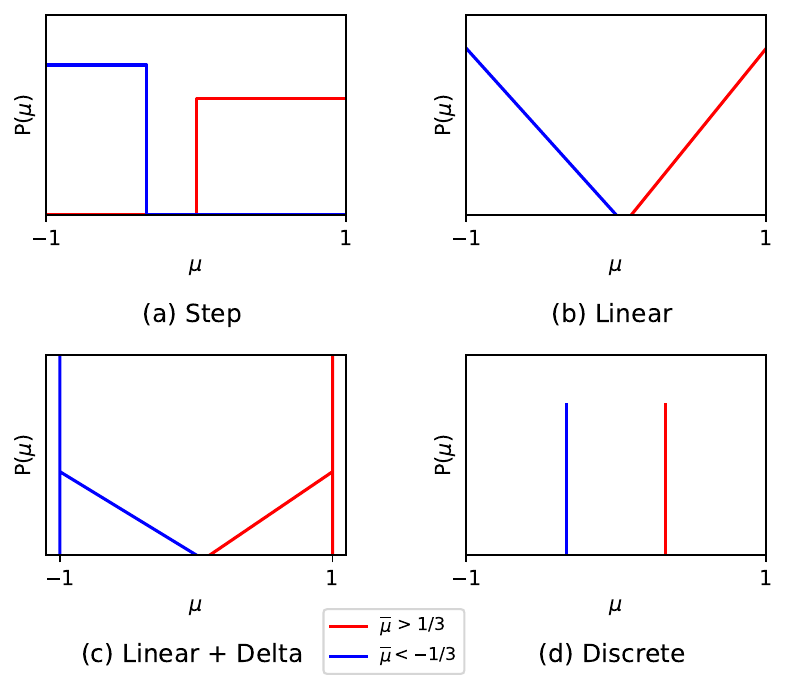}
    \caption{Shapes of the approximated angular distributions proposed by Brown and Barnett to avoid negative weights \cite{brown2008}.}
    \label{fig:approximated_pdf_reconstruction}
\end{figure}

Brown and Barnett, in their previous work, have tried to simulate problems 37 and 71, which are both linearly anisotropic, using various approximated distributions that avoided the problem of negative scattering PDFs. Their approximated distributions are reconstructed from the true angular distribution by preserving the first Legendre moment, $\bar{\mu}$, of the true distribution \cite{brown2008}. The four approximated distributions used in their work are step, linear, linear $+$ delta, and discrete. With linearly anisotropic scattering, the distribution will have negative regions if  $\abs{\bar{\mu}} > 1/3$. Figure~\ref{fig:approximated_pdf_reconstruction} gives the shape of these four approximate distributions for positive and negative $\bar{\mu}$. They are all positive everywhere, therefore, standard sampling techniques can easily be applied. We have also compared our importance sampling methodology to results obtained using these approximate distributions as another point of reference.

\subsection{Problem-37: Infinite Cylinder}
Problem-37 is a one-group infinite cylinder with a radius of $\SI{13.880411336}{\centi\meter}$ \cite{sood2003}. The cylinder's total cross section is $\SI{0.32640}{\per\centi\meter}$, capture cross section is $\SI{0.013056}{\per\centi\meter}$, fission cross section is $\SI{0.065280}{\per\centi\meter}$, and the average number of the neutrons released per fission is $2.7$. The differential scattering cross section is
\begin{equation}
   \frac{\partial\Es}{\partial\mu} = 0.248064 \left(0.5 + 1.282895 \mu\right).
\end{equation}
The method of characteristics solver in Scarabée can only be applied to problems with rectilinear outer boundaries. This limitation is overcome for this problem by placing the cylinder inside a square which has vacuum boundary conditions on all sides. The area between the cylinder and the square boundaries was then filled with a fictitious material to mimic vacuum, which had a purely absorbing cross section of $10^{-20}\si{\per\centi\meter}$. This maintains an effective vacuum boundary condition for the cylinder, as neutrons are never able to return and there is no source outside the cylinder. The geometry of the cylinder itself was approximated using a square mesh where the flat source regions had a width equal to $1/50th$ of the radius. A simple convergence study indicated that this level of geometric discretization was able to reproduce the known eigenvalue. The MOC solver in Scarabée was set to use 128 azimuthal angles with a track-spacing of $\SI{0.05}{\centi\meter}$ in conjunction with the 6 point optimal polar quadrature of Yamamoto and Tabuchi \cite{Yamamoto_tabuchi_polar_quadrature}. The convergence tolerance for $\keff$ and flux was set to $10^{-5}$.  

\begin{figure}[!hbt]
    \centering
    \includegraphics[width=1.\linewidth]{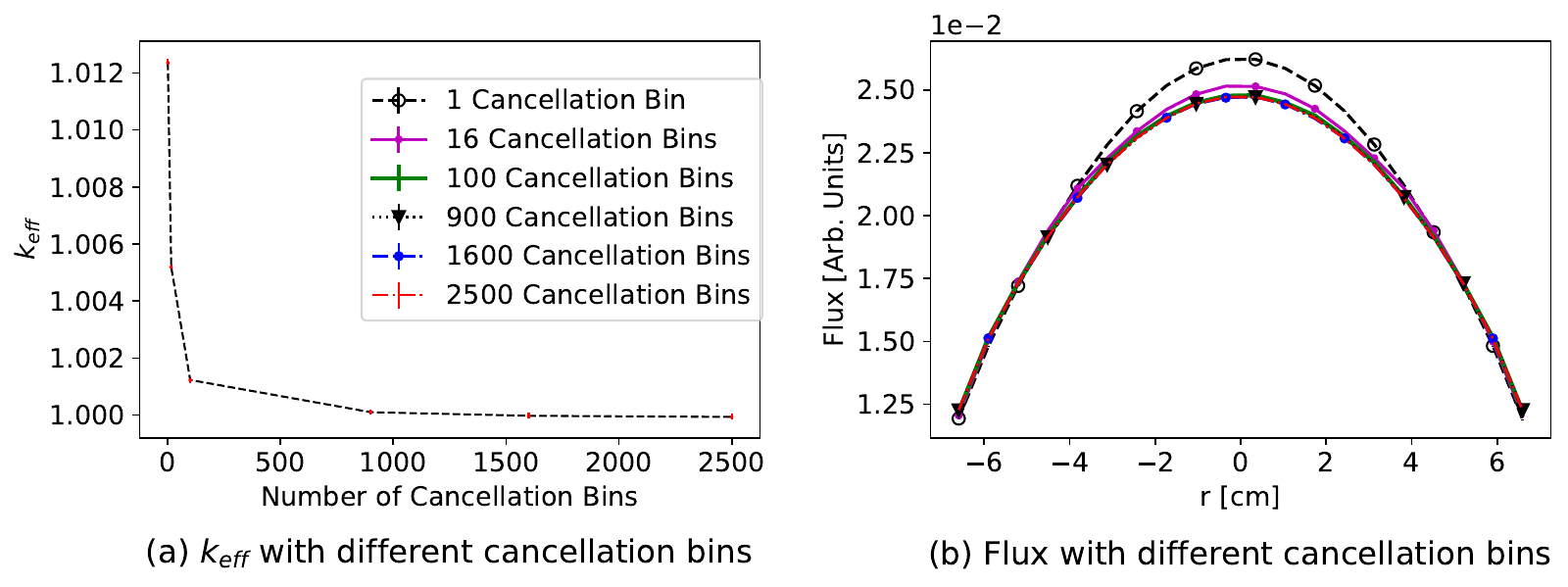}
    \caption{Cancellation mesh convergence study for problem-37}
    \label{fig:problem-37-cancellation-converegence}
\end{figure}

In the Monte Carlo simulations, the exact cylindrical geometry was used, instead of the discretization applied in Scarabée. The same fictitious absorber material was applied outside of the cylinder for the comparison of flux profiles. Simulations with different angular distributions were performed with 3 million initial source particles, with 200 inactive generations and 6000 active generations. A brief convergence study was conducted to determine the appropriate cancellation mesh for the Monte Carlo simulation using importance sampling. A regular 2D rectilinear mesh was imposed on top of the cylinder and was systematically refined until the expected value for $\keff$ was obtained. Fig.~\ref{fig:problem-37-cancellation-converegence} shows the convergence of $\keff$ and the spatial flux along the diameter of the cylinder. The spatial flux is converged with a $10 \times 10$ cancellation mesh, however $\keff$ does not appear converged until a $40 \times 40$ cancellation mesh is applied. Therefore, the $40 \times 40$ cancellation mesh was used to obtain the importance sampling results presented here. 

The estimated values of $\keff$ for all simulation methods are given in Table \ref{tab:problem-37-keff-comparision}. 
\begin{table}[!htbp] 
    \centering
    \caption{Estimated values of $\keff$ from Scarabée and Abeille for each angular distribution representation used in Problem-37}
    \begin{tabular}{lc}\toprule
         Exact & $1.000000$ \\
         \hline
         Scarabée & $1.000027$ \\
         \hline
         Step & $1.018789 \, \pm\, 0.5 \, \pcm$
         \\
         Linear & $1.018842\, \pm\, 0.5 \, \pcm $
         \\
         Linear + Delta & $1.022688\, \pm\, 0.5 \, \pcm$
         \\
         Discrete & $1.018721\, \pm\, 0.5 \, \pcm$
         \\
         \hline
         Importance Sampling & $0.999936 \, \pm 10.2 \, \pcm$
         \\
         \bottomrule
    \end{tabular}    
    \label{tab:problem-37-keff-comparision}
\end{table}
None of the approximate angular distributions predicted a $\keff$ estimate that agreed with the exact $\keff$. Their biases ranged from 3573 to 4368 standard deviations off from the known eigenvalue. The $\keff$ estimate obtained with importance sampling and weight cancellation is in excellent agreement with the exact $\keff$ to within one standard deviation. 

\begin{figure}
    \centering
    \includegraphics[width=1.\linewidth]{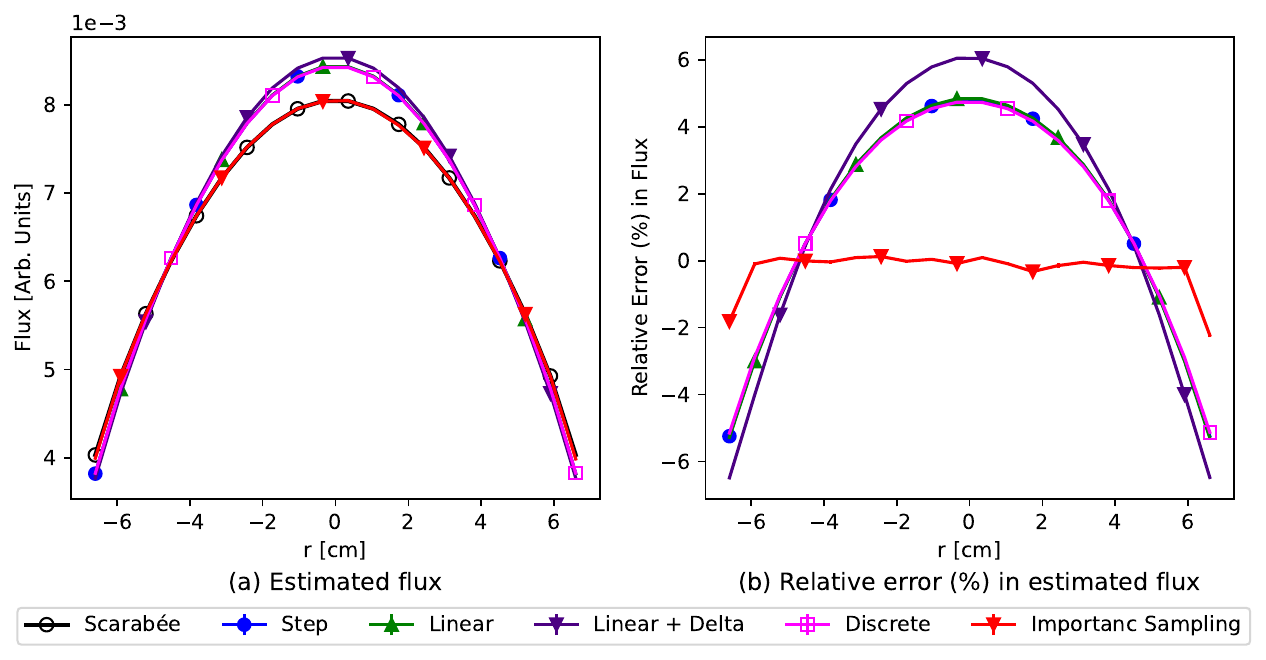}
    \caption{Estimated flux and \% relative error in the flux for each angular distribution representation in problem-37}
    \label{fig:problem-37-flux-comparision}
\end{figure}

The estimated spatial flux from each method is shown in Fig.~\ref{fig:problem-37-flux-comparision}(a), with the relative difference compared to the Scarabée solution in Fig.~\ref{fig:problem-37-flux-comparision}(b), calculated using:
\begin{equation} \label{eq:relative-error}
    \text{relative error (\%) in flux} = 100 \left( \frac{\text{flux} - \text{flux}_{\text{anisotropic scattering (MOC) }}}{\text{flux}_{\text{anisotropic scattering (MOC)}}} \right).
\end{equation}
The flux estimates obtained using approximate distributions demonstrate large errors near the center and edges of the cylinder. The linear + delta and other approximate distributions have a $~6.\%$ and $~4.8\%$ relative error, respectively, at the center of the cylinder. At the edge of the cylinder they have a $~-6.5\%$ and $~-5.3\%$ relative error, respectively. Importance sampling, however, generates results that are in near perfect agreement with the Scarabée solution. Near the center of the cylinder, the relative error does not exceed $~0.12\%$; at the edges of the cylinder, this error jumps down to nearly $-2.2\%$. This increase in the error is not necessarily problematic, as the method of characteristics estimation of the flux in this region is likely less accurate due to the effective vacuum boundary, finite number of angles considered, and the discretization of the cylinder geometry into a rectilinear mesh. Due to these effects, it is likely that the flux from Scarabée is less accurate than the Monte Carlo in these regions, rather than the Monte Carlo importance sampling algorithm deteriorating in these regions. 

\subsection{Problem-71: Infinite Slab}
\begin{figure}[!hbt]
    \centering
    \includegraphics[width=1.\linewidth]{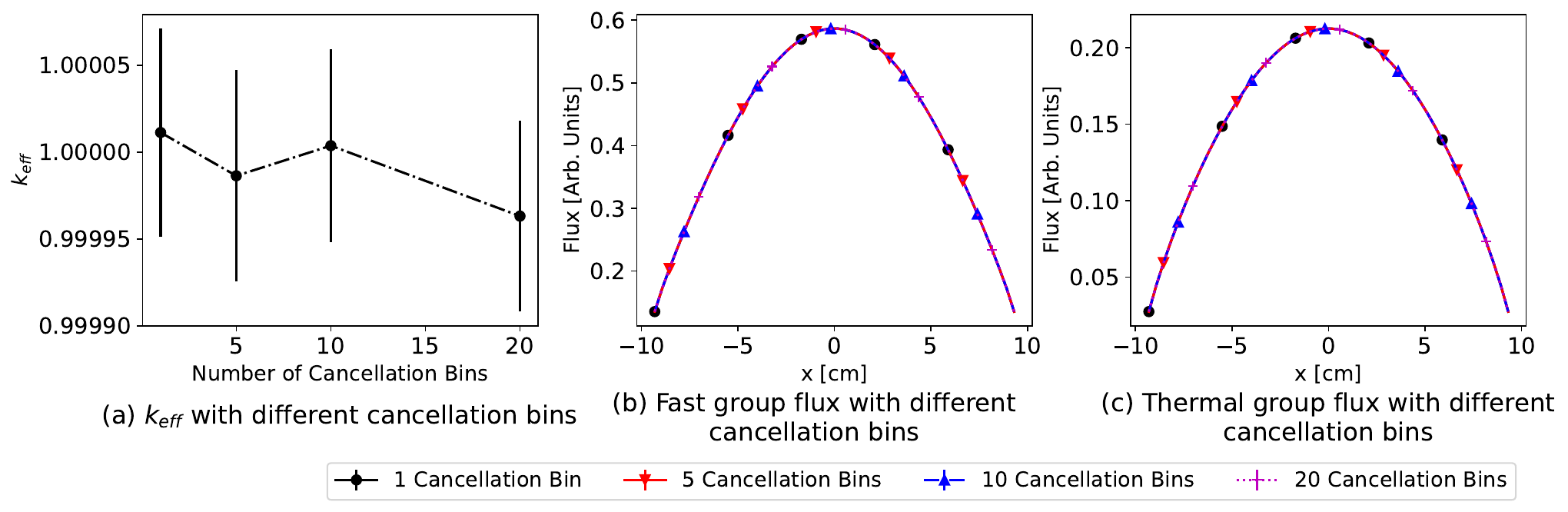}
    \caption{Cancellation mesh convergence study for problem-71.}
    \label{fig:problem-71-cancellation-convergence}
\end{figure}

Problem-71 is a two-group infinite slab with a thickness of $\SI{18.9918}{\centi\meter}$ \cite{sood2003}. The macroscopic cross-sections for the problem are given in Table ~\ref{tab:problem-71-cross-sections}.
\begin{table*}[!htbp]
    \centering
    \caption{Macroscopic cross sections for problem-71}
    \begin{tabular}{lccccc}\toprule
         Group & $\nu$ & $\Sigma_f$ & $\Sigma_c$ & $\Sigma_t$ & $\chi$ \\
         \midrule
         1 (Fast) & 2.5 & 0.0010484 & 0.0010046 & 0.65696 & 1. \\
         2 (Thermal) & 2.5 & 0.050632 & 0.025788  & 2.52025 & 0. \\
         \bottomrule
         \\
    \end{tabular}
    
    \begin{tabular}{lcc}
         \toprule
         Group (g) & $\partial\Sigma_{s, g \rightarrow 1}(\mu)/\partial\mu$ & $\partial\Sigma_{s, g \rightarrow 2}(\mu)/\partial\mu$
         \\
         \midrule
         1 & $0.62568 \, ( 0.5 + 0.6582997698\,\mu )$ & $ 0.029227 \, ( 0.5 + 0.2591336778\,\mu )$ 
         \\
         2 & 0. & $ 2.44383\,( 0.5 + 0.3409320616\,\mu )$ 
         \\
         \bottomrule
    \end{tabular}
    \label{tab:problem-71-cross-sections}
\end{table*}
The use of approximated angular distributions is only employed for those group-to-group energy transfers which have negative regions in the scattering distribution. In this problem, negative regions are present in the group-1-to-1 and group-2-to-2 energy transfers. The Monte Carlo simulations for this problem were performed with 1 million initial source particles, with 500 inactive generations and 5000 active generations. Fig.~\ref{fig:problem-71-cancellation-convergence} shows the convergence study of the cancellation mesh used for importance sampling. The simulation could be said to be well converged with 1 cancellation bin. Unlike in problem-37, here, $\keff$ does not change much with further refinements of the cancellation mesh as shown in Fig.~\ref{fig:problem-71-cancellation-convergence}(a). The spatial flux along the slab also appears converged with a single cancellation bin, as apparent in Fig.~\ref{fig:problem-71-cancellation-convergence}(b) and (c). Though the results were well converged with only 1 cancellation bin, it was decided to use the 20-bin cancellation mesh out of an abundance of caution, to ensure there there was no bias from cancellation.

\begin{table}[!htp] 
    \centering
    \caption{Estimated values of $\keff$ from Scarabée and all Monte Carlo angular distribution representations for Problem-71.}
    \begin{tabular}{lc}\toprule
         Exact & $1.000000$ \\
         \hline
         Scarabée & $0.999992$ \\
         \hline
         Step & $0.999742 \, \pm\, 1.4 \, \pcm$
         \\
         Linear & $1.000218\, \pm\, 1.4 \, \pcm $
         \\
         Linear + Delta & $1.001018\, \pm\, 1.4 \, \pcm$
         \\
         Discrete & $0.999164\, \pm\, 1.3 \, \pcm$
         \\
         \hline
         Importance Sampling & $ 0.999963 \, \pm \, 5.5 \, \pcm $
         \\
         \bottomrule
    \end{tabular}    
    \label{tab:problem-71-keff-comparision}
\end{table}
All estimated $\keff$ values for problem-71 are given in Table~\ref{tab:problem-71-keff-comparision}. The analytically known eigenvalue for this problem is again unity. None of the Monte Carlo simulations using approximate distributions produced a multiplication factor that agreed with unity within one standard deviation. For these approximate distributions, the biases in the $\keff$ estimations range from 16 to 73 standard deviations. Importance sampling, however, produces an estimate which is in excellent agreement with the the analytical known $\keff=1$ to within one standard deviation.

\begin{figure}[!h] 
    \centering
    \includegraphics[width=1.\linewidth]{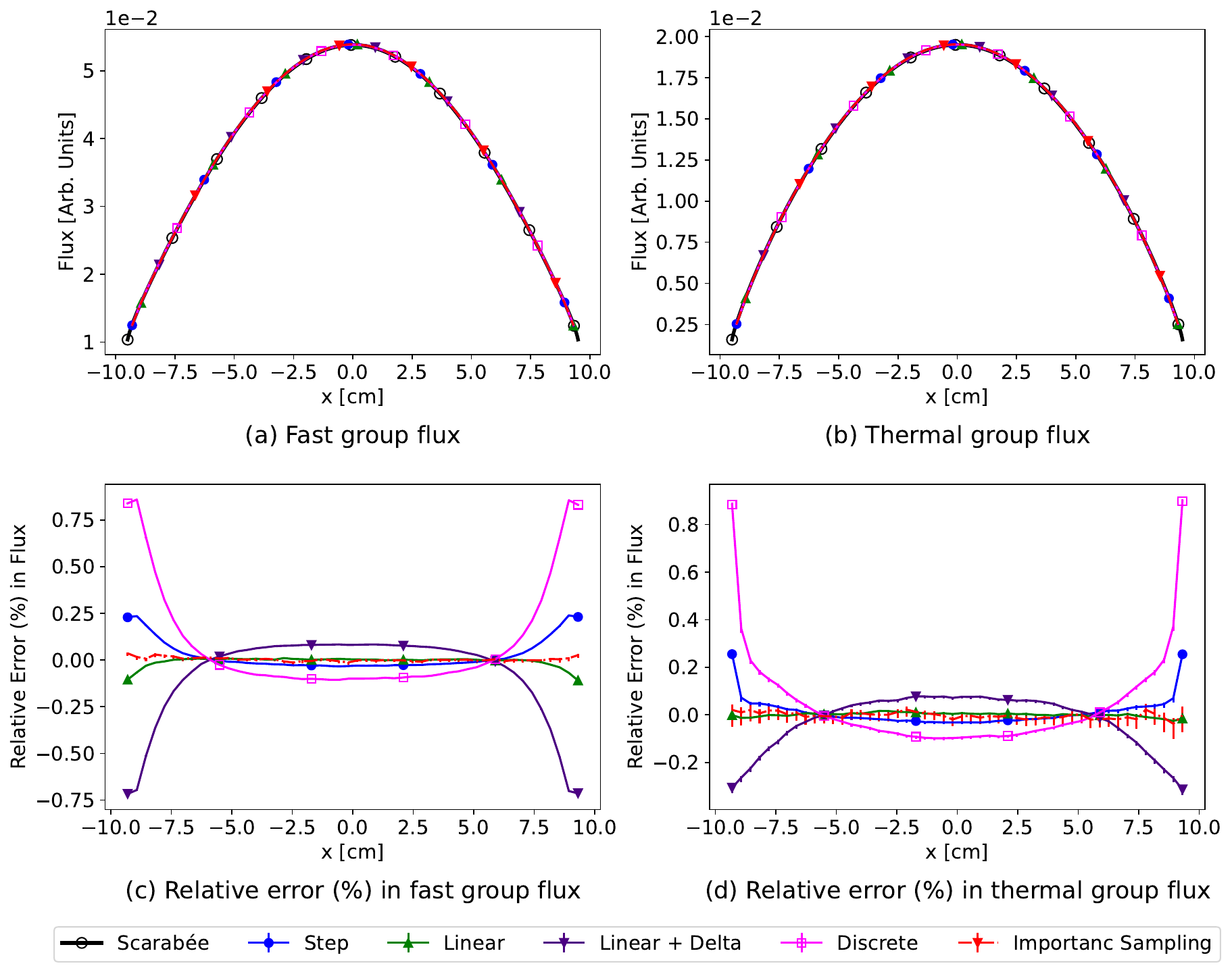}
    \caption{Estimated flux and \% relative error for all Monte Carlo simulation compared to Scarabée for problem-71.}
    \label{fig:problem-71-flux-comparision}
\end{figure}
Fig.~\ref{fig:problem-71-flux-comparision} provides a comparison of the estimated flux for each angular distribution representation in each group, and the \% relative error compared to the reference Scarabée solution. The relative error in flux with the linear + delta approximate distribution goes up to as much as $-0.72\%$ and $-0.31\%$ in the fast and thermal groups, respectively, at the edges of the slabs. The discrete distribution also has a significant relative error at the edge, going up to $0.9\%$ in both groups. The two other approximate distributions had relative errors between $-0.1\% \text{ to } 0.26\%$ at the slab edges in both groups. All approximate distributions had relative errors between  $-0.1\% \text{ to } 0.8\%$ at the center of the slab. The flux estimates from importance sampling had relative errors with the Scarabée solution that were between $-0.04\% \text{ to } 0.04\%$. One might notice in Figures~\ref{fig:problem-71-flux-comparision}(c) and (d) that the importance sampling results have error bars that are noticeably larger than the approximate distributions. This is explained by the use of negative weights which is increasing the statistical variance in the scores for the flux.

These two simple example problems demonstrate that approximated angular distributions can be very inefficient at computing accurate values for $\keff$ and the flux. It is also difficult to predict or understand which of these approximated distributions will yield a more accurate results for a given problem. However, using importance sampling with weight cancellation, it is clearly possible to use the truncated scattering distribution with negative regions in Monte Carlo simulations and obtain reference results. These test problems were admittedly quite simple, however, and do not represent realistic light water reactor simulations. In the next section, our importance sampling with weight cancellation method is applied to a real-world research reactor.

\section{Research Reactor Model}\label{sec:RCF}
The RPI Research Critical Facility (RCF) is a zero power light-water reactor. Being comprised of a small 333 fuel pin configuration in an open pool, it is fairly sensitive to core leakage, and therefore the scattering angular distributions. This makes the RCF an ideal case for studying the application of importance sampling with approximate regional cancellation to treat anisotropic scattering in multigroup Monte Carlo simulations of reactors. We will briefly outline the geometry used in this study and cover how multigroup cross sections were generated for the problem. Simulation results with importance sampling are then presented in Sec.~\ref{sec:rcf_sims}.

\subsection{Geometry Description}\label{sec:rcf_geom}
\begin{figure}
    \centering
    \includegraphics[width=0.5\linewidth]{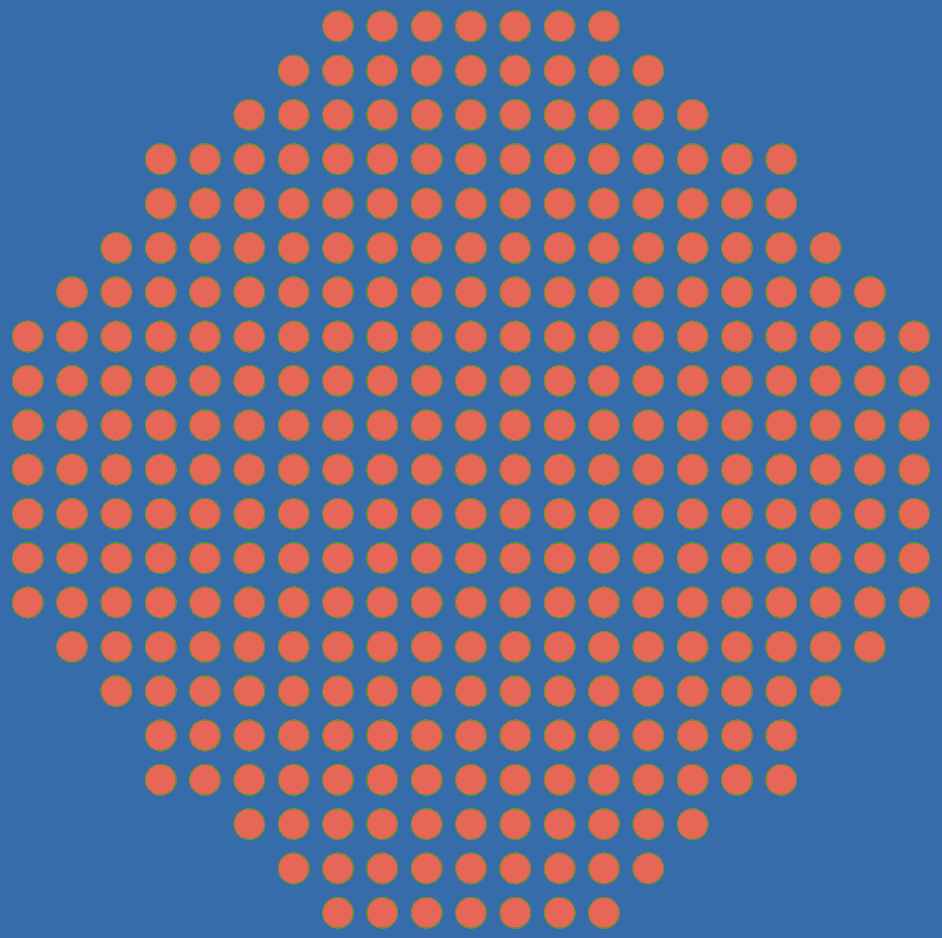}
    \caption{Fuel lattice layout of the RCF.}
    \label{fig:RCF-lattice}
\end{figure}

In the present study, a 2D variant of the RCF is considered, to facilitate its simulation with the Scarabée MOC solver for obtaining reference solutions with anisotropic scattering. The RCF uses \ce{UO2} fuel that is $ 4.81 $ weight percent enriched with a pellet radius of $\SI{0.53340}{\centi\meter}$. The gas between the fuel pellet and cladding is filled with a helium-hydrogen mixture having a thickness of $\SI{0.00762}{\centi\meter}$. The thickness of the stainless steel cladding is $\SI{0.05026}{\centi\meter}$. The fuel pin lattice arrangement is shown in the Fig.~\ref{fig:RCF-lattice}. There are 333 fuel pins forming a symmetric lattice structure, with a pin pitch of $\SI{1.6256}{\centi\meter}$. The fuel lattice sits at the center of a large stainless-steel tank, which is filled with water when in operation. The inner radius of the tank is $\SI{106.68}{\centi\meter}$ and the thickness is $\SI{1.12}{\centi\meter}$. Further details and pictures on the RCF can be found in other publications \cite{Dupont2019,Dupont2023}.

\begin{figure}[!hbt]
    \centering
    \includegraphics[width=1.\linewidth]{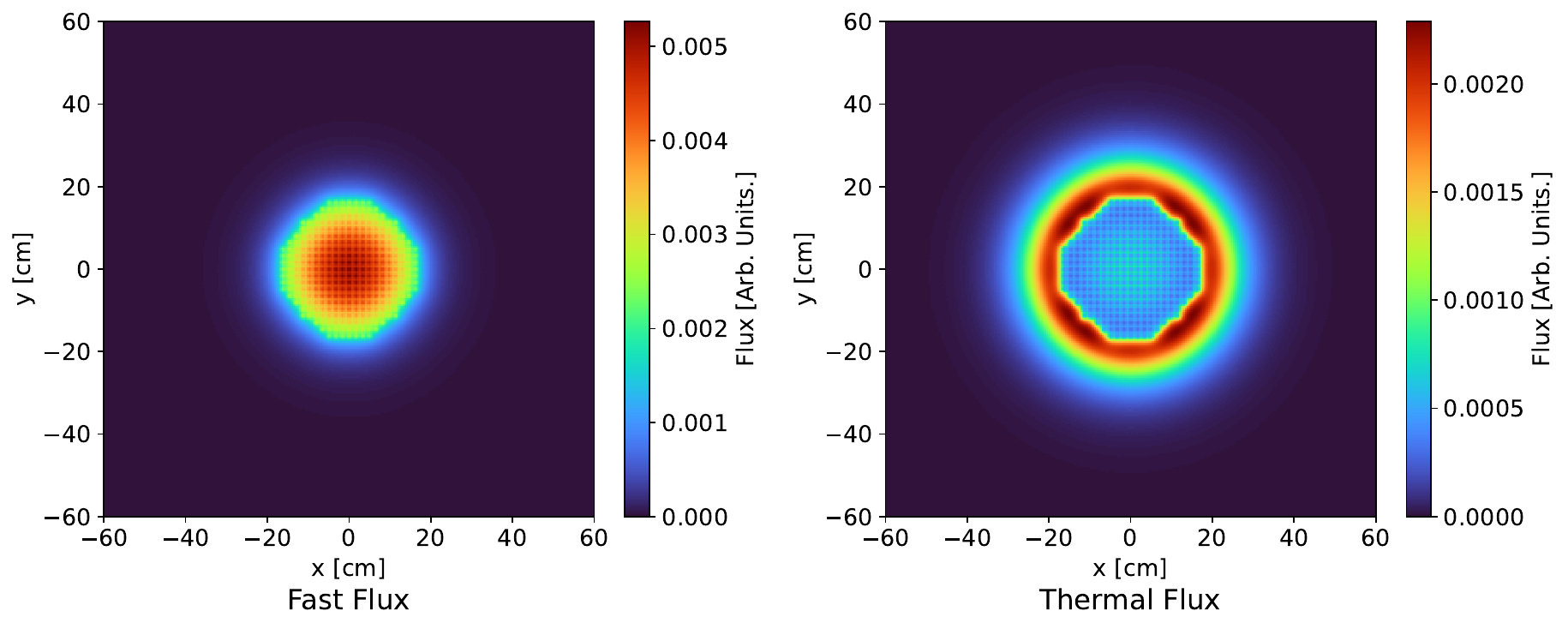}
    \caption{Fast and thermal scalar flux obtained from a continuous energy simulation using the \textit{Abeille} Monte Carlo code.}
    \label{fig:RCF-continuous-fast-thermal-scalar-flux-plots}
\end{figure}

\begin{figure}[!hbt]
    \centering
    \includegraphics[width=0.7\linewidth]{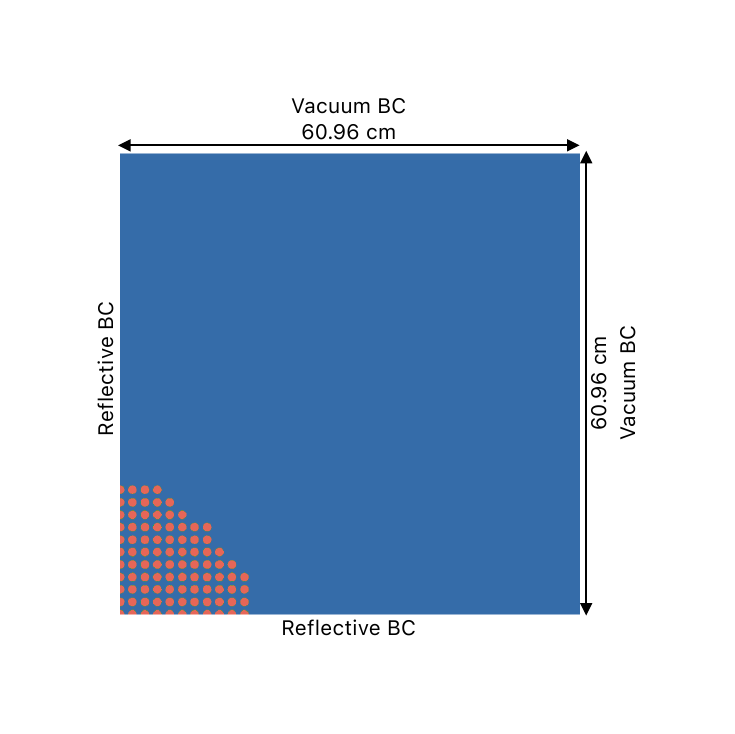}
    \caption{Simplified RCF geometry, used in all multigroup simulations.}
    \label{fig:RCF-modified-geometry}
\end{figure}

The Fig.~\ref{fig:RCF-continuous-fast-thermal-scalar-flux-plots} shows the fast and thermal fluxes, obtained with the Abeille Monte Carlo Code in continuous energy. From this figure, it is evident that the RCF's water tank is effectively an infinite reflector, and there is negligible flux beyond a $\SI{60}{\centi\meter}$ radius from the center of the core. As such, we have chosen to simplify the geometry in our multigroup simulations by applying vacuum boundary conditions at $x=\pm\SI{60.96}{\centi\meter}$ and $y=\pm\SI{60.96}{\centi\meter}$, ignoring the outer reflector region and tank wall. We also have chosen to take advantage of the symmetry of the core and to apply reflective boundary conditions along the $x$ and $y$ axes, simulating only a $1/4$ core in 2D. These simplifications greatly reduce the simulation run time and memory requirements when using Scarabée for the MOC solutions. The final simulated RCF geometry is shown in Fig.~\ref{fig:RCF-modified-geometry}.

\subsection{Multigroup Cross Section Generation}\label{sec:rcf_xs_gen}
The OpenMC continuous energy Monte Carlo code was used to generate the multigroup cross sections for the problem \cite{Romano2015,Romano2024}. A unique set of cross sections was generated for the fuel, gas-gap, cladding, and water. The fuel and cladding cross sections were homogenized across all instances of those materials in the core, neglecting variations in the flux spectrum at different fuel positions. A 33 group structure was used, and the upper energy bound of each group is given in Table~\ref{tab:33-group-structure}. Scattering matrices up to the $L=3$ Legendre order were tallied in the simulation. The OpenMC simulation was performed with $10^6$ particles and 4000 active generations; 200 inactive generations are taken to allow for fission source convergence. 
\begin{table}
    \caption{Upper energies of the groups in the 33 group-structure. }
    \centering
    \begin{tabular}{cl|cl|cl}
         Group & Energy (MeV) & Group & Energy (MeV) & Group & Energy (MeV)
         \\
         \hline
         $1$ & $19.6403$ & $12$ & $6.737938E-02$ & $23$ & $2.837502E-04$
         \\
         $2$ & $9.999987$ & $13$ & $4.086766E-02$ & $24$ & $1.327005E-04$
         \\
         $3$ & $6.065299$ & $14$ & $2.499908E-02$  & $25$ & $8.895177E-05$
         \\
         $4$ & $3.328707$ & $15$ & $1.489967E-02$ & $26$ & $6.144204E-05$
         \\
         $5$ & $2.231299$ & $16$ & $9.118808E-03$ & $27$ & $4.016895E-05$
         \\
         $6$ & $1.336941$ & $17$ & $5.004508E-03$ & $28$ & $2.253556E-05$
         \\
         $7$ & $8.600058E-01$ & $18$ & $3.481068E-03$ & $29$ & $1.354604E-05$
         \\
         $8$ & $4.940018E-01$ & $19$ & $1.910451E-03$ & $30$ & $8.300322E-06$
         \\
         $9$ & $3.206464E-01$ & $20$ & $1.135007E-03$ & $31$ & $4.000000E-06$
         \\
         $10$ & $1.950077E-01$ &  $21$ & $7.485173E-04$ & $32$ & $5.200108E-07$
         \\
         $11$ & $1.156235E-01$ & $22$ & $4.107950E-04$ & $33$ & $1.042977E-07$
         \\
                &       &           &               &   & $1.100027E-10$
         \\
    \end{tabular}
    \label{tab:33-group-structure}
\end{table}

The transport correction is often used in multigroup simulations to approximate the effects of anisotropic scattering. This takes the form of a modification to the total cross section and the P0 scattering matrix which are thereafter referred to as the transport cross section and the transport scattering matrix, respectively. Using these transport corrected cross sections, it is then assumed that all scattering is isotropic. The most common convention for obtaining transport corrected cross sections is the outflow approximation where the transport cross section for group $g$ can be obtained with
\begin{equation}{\label{Etr_outflow_trxs}}
    \Sigma_{tr,g} = \Sigma_{t,g} - \sum_{g'=0}^{G}\Sigma_{s,1,g\rightarrow g'} = \Sigma_{t,g} - \bar{\mu}\Sigma_{s,g},
\end{equation}
and the transport corrected scattering matrix can be obtained with
\begin{equation}{\label{Estr_outflow_trxs}}
    \Sigma_{str,g'\rightarrow g} = \Sigma_{s,0,g'\rightarrow g} - \delta_{g',g} \sum_{g'=0}^{G}\Sigma_{s,1,g\rightarrow g'} = \Sigma_{s,0,g'\rightarrow g} - \delta_{g',g}\bar{\mu}\Sigma_{s,g},
\end{equation}
where, $\delta_{g,g'} $ is Kronecker delta function\cite{Stammler1983}.

While the outflow approximation is very common in the realm of reactor analysis, previous literature has indicated that it cannot adequately account for all anisotropic scattering effects, particularly in the case of light water \cite{Imapact_of_inflow_tr}. To avoid this error, many have suggested the use of the more accurate (though more difficult to implement) inflow approximation for water \cite{Imapact_of_inflow_tr, Smith2017}. Choi et al.\ demonstrated on the B\&W1484 I core that the outflow approximation results in a bias in $\keff$ of more than $1220 \pcm$, while the inflow approximation on the moderator only had a bias of $25 \pcm$ in 2D and $41 \pcm$ in 3D, compared to reference solutions \cite{Imapact_of_inflow_tr}. Given the large effect that the transport correction approximation of the moderator has on the results, we have also chosen to compare our explicit anisotropic scattering against the outflow transport correction and the inflow transport correction (applied only to the moderator). The inflow approximation requires the P1 flux moments, $\phi_{1,g}$. Assuming these are available, one can compute the transport cross section and transport corrected scattering matrix with
\begin{equation}{\label{eq:Etr_inflow_trxs}}
    \Sigma_{tr,g} = \Sigma_{t,g} - \frac{\displaystyle\sum_{g'=0}^{G}\Sigma_{s,1,g'\rightarrow g}\phi_{1, g'}}{\phi_{1, g}}
    \text{ and}
\end{equation}
\begin{equation}{\label{eq:Estr_inflow_trxs}}
    \Sigma'_{s,g'\rightarrow g} = \Sigma_{s,0,g'\rightarrow g} - \delta_{g',g} \frac{\displaystyle\sum_{g'=0}^{G}\Sigma_{s,1,g'\rightarrow g}\phi_{1, g'}}{\phi_{1, g}}.
\end{equation}

To compute the inflow transport correction for the moderator, a method similar to that proposed by Herman was employed \cite{HermanPhD}. The homogeneous P1 equations were solved for water using the fission spectrum from $\isotope[235]{U}$ as a source and the computational method outlined by Stamm'ler and Abbate \cite{Stammler1983}. This results in the group-wise diffusion coefficients, $D_g$, from which the group-wise transport cross sections can be determined as
\begin{equation}
    \Sigma_{tr,g} = \frac{1}{3 D_g}.
\end{equation}
Subsequently, the transport correction for each group can be determined as
\begin{equation}
    \frac{\displaystyle\sum_{g'=0}^{G}\Sigma_{s,1,g'\rightarrow g}\phi_{1, g'}}{\phi_{1, g}} = \Sigma_{t,g} - \Sigma_{tr,g}.
\end{equation}
Once known, this correction can then be used to generate the transport scattering matrix. It is important to emphasize that in our simulations the inflow approximation \emph{is only used on the moderator}, while all other materials in the simulation still employ the outflow approximation for the transport correction.

\subsection{Results}\label{sec:rcf_sims}
The Scarabée MOC simulation of the RCF requires a geometric discretization for the flat source regions. Each fuel cell was discretized into 8 angular divisions. Subsequently, the fuel pellet was divided into two annular regions with equal areas. One ring was used for the gas-gap and the cladding. The water moderator surrounding the pin is also divided into two annular regions with the first having a diameter equal to the pin pitch and second extending from this ring out to the boundary of the pin cell. 

As previously seen in Fig.~\ref{fig:RCF-continuous-fast-thermal-scalar-flux-plots}, the gradient of the flux is quite large in the reflector close to the fuel lattice. Therefore, a fine mesh is needed to adequately represent the scalar flux with flat source regions. The region $0 \leq x \leq 43.0784 \cm$ and $0 \leq y \leq 43.0784 \cm$ of the reflector was divided into square regions with a width of $1/12$th of the pin pitch ($\SI{0.1354667}{\centi\meter}$). Beyond this domain, the width is increased to $1/4$th of the pin pitch ($\SI{0.4064}{\centi\meter}$). This approach greatly reduces the memory consumption and computational time while also maintaining adequate fidelity in the results. The MOC solver in Scarabée was set to use 128 azimuthal angles with a track-spacing of $\SI{0.01}{\centi\meter}$ and the 6 point optimized polar quadrature of Yamamoto and Tabuchi \cite{Yamamoto_tabuchi_polar_quadrature}. The convergence tolerance for $\keff$ and the flux was set to $10^{-5}$. For this study, because we are evaluating a new method for treating anisotropic scattering in multigroup Monte Carlo, the reference solution is taken to be the Scarabée results obtained when using explicit anisotropic scattering based on the spherical harmonics expansion. The Monte Carlo results, as well as the other Scarabée results obtained with the outflow and inflow transport corrections, are compared to the reference anisotropic scattering Scarabée solution.

The mutligroup Monte Carlo results with importance sampling were again obtained with the Abeille Monte Carlo code. The simulation was performed with $10^7$ initial source particles, with 200 inactive generations and 2000 active generations. The delta-tracking transport algorithm was employed \cite{lux_koblinger_MC_book_1991} and collisions were treated using a branchless algorithm \cite{Belanger2023}. All material cross sections used scattering matrices up to the $L=3$ Legendre order, and all materials therefore had some negative angular distributions for certain group-to-group transfers. For most of the groups with negative PDFs, the likelihood ratio given by Eq.~\eqref{eq:weight_modifier_ratio} was between $1$ and $2$, however, there were some angular distributions where the ratio exceeded $2$. The largest value of the likelihood ratio was $2.60327$ when up-scattering from group-20 to group-19 in water (this energy transition has a very low probability of being sampled). If absolute value of a particle's weight exceeded 2, the particle was split in the simulation. 

\begin{figure}[!hbt]
    \centering
    \includegraphics[width=1.\linewidth]{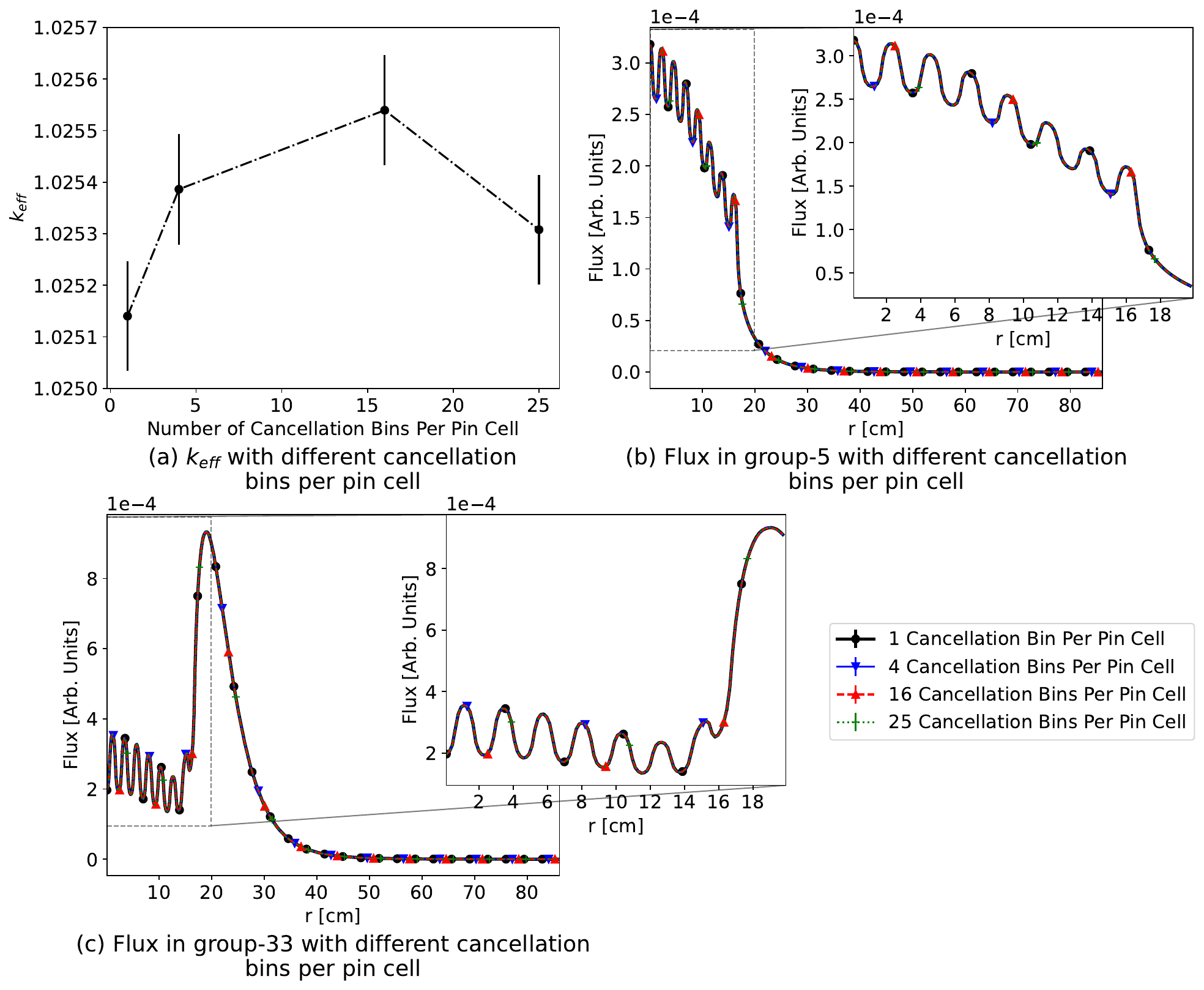}
    \caption{Cancellation mesh convergence study for the Monte Carlo simulation using importance sampling and weight cancellation.}
    \label{fig:RCF-cancellation-study}
\end{figure}
To determine an appropriate cancellation mesh, a convergence study was performed on the simplified RCF geometry. Fig.~\ref{fig:RCF-cancellation-study} shows the results of this study. Four different mesh sizes were considered, each one subdividing a pin cell in a different manner: $1 \times 1$, $2\times2$, $4 \times 4$, and $5 \times 5$. Fig.~\ref{fig:RCF-cancellation-study}(a) shows the variation of $\keff$ with different cancellation meshes. From this plot, it could be said that $\keff$ is already converged with only 1 cancellation bin per pin cell, as refining the cancellation mesh leads to only a minimal change in $\keff$, and all estimates of $\keff$ for the different mesh refinements are in agreement to within two standard deviations. Fig.~\ref{fig:RCF-cancellation-study}(b) and (c) show the flux along the diagonal of the geometry in groups 5 and 33 respectively, obtained using different cancellation meshes. Upon close inspection, the flux in these groups does not appear to change with refinement of the cancellation mesh. While it appears that having only 1 cancellation bin per pin cell is adequate in this case, we decided to use $5\times 5$ cancellation bins per pin cell out of an abundance of caution. Future work will further explore this interesting observation.

\begin{table}[!hbt]
    \caption{Estimates of $\keff$ for the simplified RCF system.}
    \centering
    \begin{tabular}{llll}
         Method & & $\keff $ & $\Delta\keff$ 
         \\
         \toprule
         Anisotropic Scattering & (MOC) & $1.026599$ & 
         \\
         Importance Sampling & (MC) & $1.025308 \pm 10.6 \, \pcm$ & $129.1 \pcm$
         \\
         Inflow Approximation & (MOC) & $1.026724$ & $12.5 \pcm$
         \\
         Outflow Approximation & (MOC) & $1.014660$ & $1193.9 \pcm$
          \\
    \end{tabular}
    \label{tab:RCF-keff-comparison-obtained-for-different-method}
\end{table}
The estimates of $\keff$ from all simulations are given in Table~\ref{tab:RCF-keff-comparison-obtained-for-different-method}. The $\keff$ obtained with Monte Carlo using importance sampling is in very good agreement with the anisotropic scattering MOC results from Scarabée. While the absolute difference between the two is approximately $130 \pcm$, it is important to recognize that the Scarabée results likely have many discretization errors due to the use of flat source regions and a finite angular discretization. The Monte Carlo results from Abeille on the other hand are perfectly continuous with respect to the space and direction variables. As will be seen later on, it is likely that the Abeille results are closer to the ``true'' value for this system. The absolute difference in $\keff$ between the anisotropic MOC simulation and the inflow transport correction MOC simulation is only $12.5 \pcm$. Both simulations are using the same spatial and directional discretizations, so there should be no bias between the two results in this regard. Therefore, we can confirm other authors' previous findings that the inflow treatment for the moderator is far superior to the outflow approximation, as the latter had a bias in $\keff$ that was approximately $1190 \pcm$. Again, since the same geometric discretizations were used, it can be inferred that the entirety of this bias is attributed to the transport correction approximation.

\begin{figure}
    \centering
    \includegraphics[width=1.\linewidth]{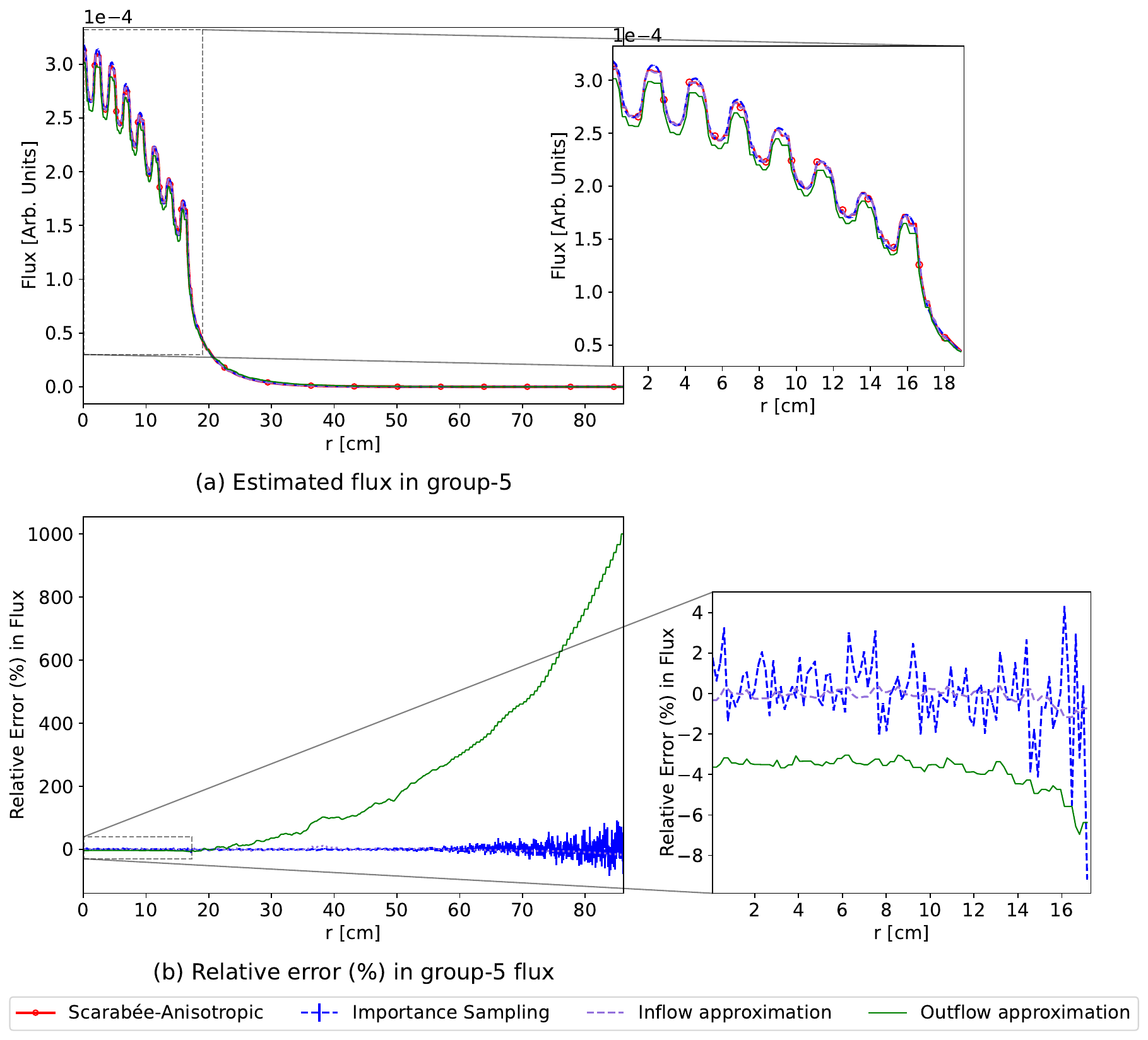}
    \caption{Group-5 estimated scalar flux and associated \% relative error along the diagonal of the modified RCF geometry by different methods }
    \label{fig:RCF-flux-comparision-group-5}
\end{figure}

The fission neutrons are born with average energy of approximately 2 MeV, which falls into group 5. For this group, the flux calculated with all four methods is shown in Fig.~\ref{fig:RCF-flux-comparision-group-5}(a), plotted along the diagonal of the geometry. The \% relative error compared to the anisotropic scattering MOC results are shown in Fig.~\ref{fig:RCF-flux-comparision-group-5}(b) calculated using Eq.~\eqref{eq:relative-error}. 
The flux near the fuel region is well predicted by importance sampling as well as the inflow approximation. In the fuel region, the relative error in importance sampling is typically less than 2\%. In the reflector, the relative error in importance sampling is between $-9 \text{ to } 4\%$. Close to the vacuum boundary condition, there is a much higher variance which is dominated by the statistical error in the Monte Carlo results, due to the small number of particles that travel that far away from the center of the core. The inflow approximation has the relative error between $-1 \text{ to } 1 \%$ in the fuel regions, and can go as high as 10\%.  In general, the inflow transport correction also has excellent agreement with the reference solution. The outflow approximation does not perform as well, however, systematically underestimating the flux in the fuel lattice region compared to the other methods. Moving away from the center of the core, the error in the outflow approximation flux increases drastically, though the flux in this region is extremely small in this region of the problem in general.
\begin{figure}
    \centering
    \includegraphics[width=1.\linewidth]{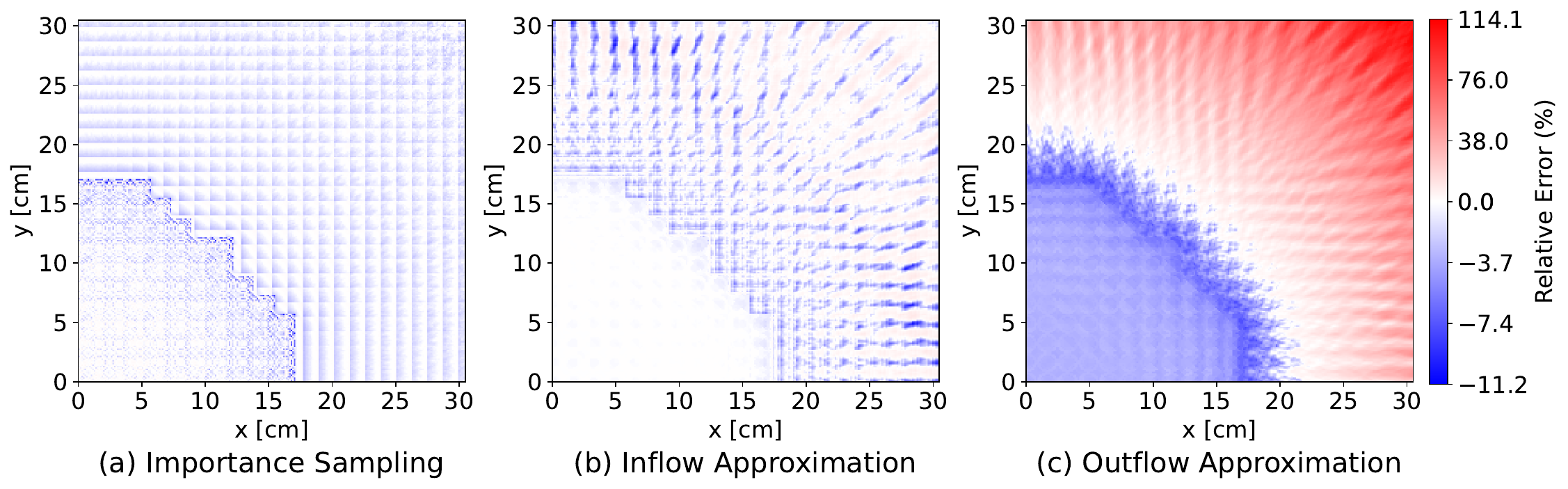}
    \caption{2D contour of relative error (\%) in scalar flux of group-5 calculated by different methods}
    \label{fig:RCF-flux-2D-contour-rel-error-group-5}
\end{figure}
Fig.~\ref{fig:RCF-flux-2D-contour-rel-error-group-5} (a), (b), and (c) provide 2D depictions of the relative error of the importance sampling, inflow, and outflow approximation fluxes, respectively, in group-5. In this group, the relative error in the flux with importance sampling is between $-11.2 \text{ to } 12.1 \%$. For the inflow approximation, the relative error is between $-10.5 \text{ to } 11.3\%$. Interestingly, close examination of Fig.~\ref{fig:RCF-flux-2D-contour-rel-error-group-5}(a) shows that the largest magnitude relative errors coincide with the boundaries of the flat source regions (particularly at the fuel assembly - reflector interface), and is an indication that the ``error'' in the Monte Carlo solution is driven by the geometric discretization in the MOC solution, and that the importance sampling method likely has more accurate results than those obtained with the anisotropic MOC solver.

\begin{figure}[!hbt]
    \centering
    \includegraphics[width=1.\linewidth]{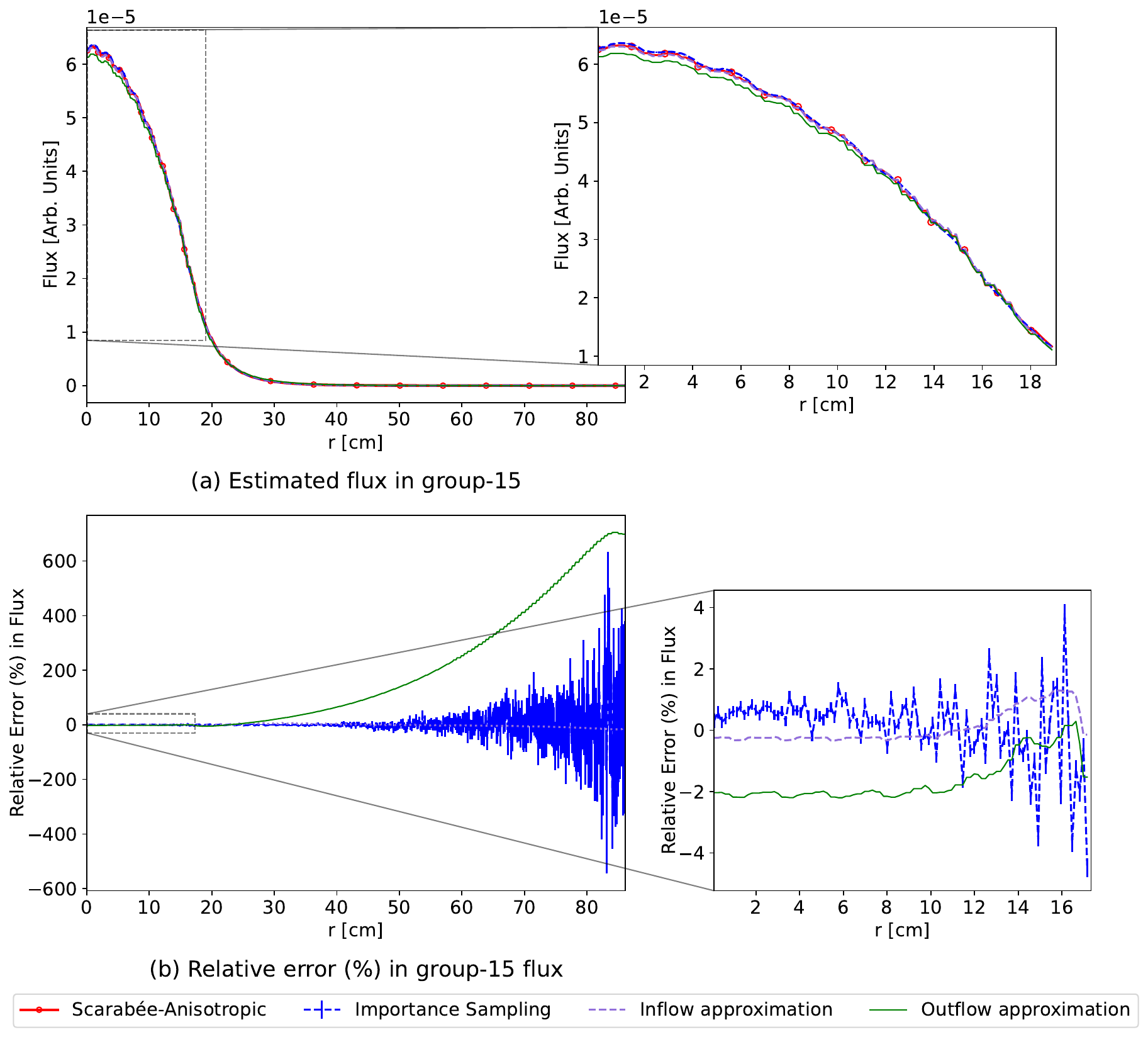}
    \caption{Group-15 estimated scalar flux and associated relative error (\%) along the diagonal of the modified RCF geometry by different methods }
    \label{fig:RCF-flux-comparision-group-15}
\end{figure}
\begin{figure}[!bt]
    \centering
    \includegraphics[width=1.\linewidth]{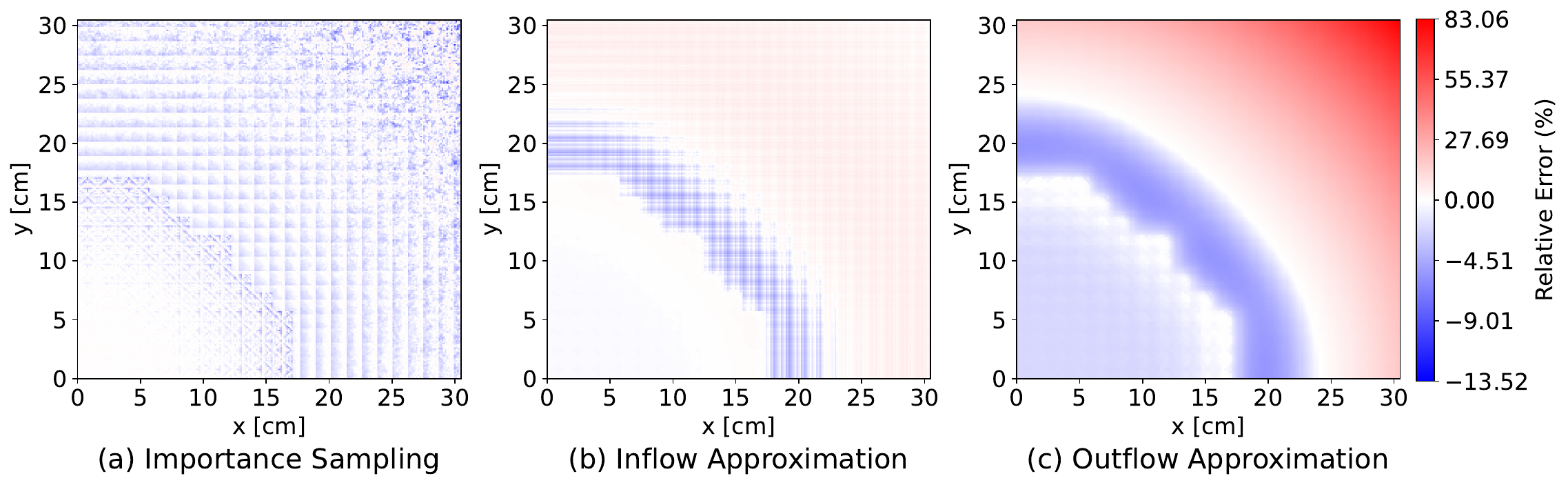}
    \caption{2D contour of relative error (\%) in scalar flux of group-15 calculated by different methods}
    \label{fig:RCF-flux-2D-contour-rel-error-group-15}
\end{figure}

Group-15 is the highest energy group which lies within the resolved resonance region of \isotope[238]{U}. Fig.~\ref{fig:RCF-flux-comparision-group-15}(a) and (b) show the scalar flux along the diagonal and the relative error in the flux, respectively. Importance sampling and the inflow approximation are again in excellent agreement with the reference anisotropic MOC results. However, deep in the reflector, the standard deviation in the relative error is quite large for importance sampling. This is likely driven by the presence of negative weights in the simulation which will increase the variance; the problem is exacerbated far from the center of the core where very few particles venture. There are even a few tally bins where the estimated average flux from the importance sampling simulation is negative, though the associated standard deviation is greater than 100\%, therefore including physically valid positive values of the flux. Fig.~\ref{fig:RCF-flux-2D-contour-rel-error-group-15}(a) shows the 2D contour of the relative error for importance sampling. While the errors closer to the fuel lattice still appear to be dominated by the flat source regions, the opposing corner of the geometry is now clearly being influenced by the statistical uncertainty of the Monte Carlo simulation, due to the low number of particles in that region and the negative statistical weights. Again, the outflow approximation results in the largest biases. Interestingly, both transport corrected approximations appear to have a similar out-in bias where the bias is positive in the outer reflector, negative in the region of the reflector close to the fuel lattice, and slightly less negative in the fuel region. This qualitative behavior is not apparent in the importance sampling simulation.

\begin{figure}[!hbt]
    \centering
    \includegraphics[width=1.\linewidth]{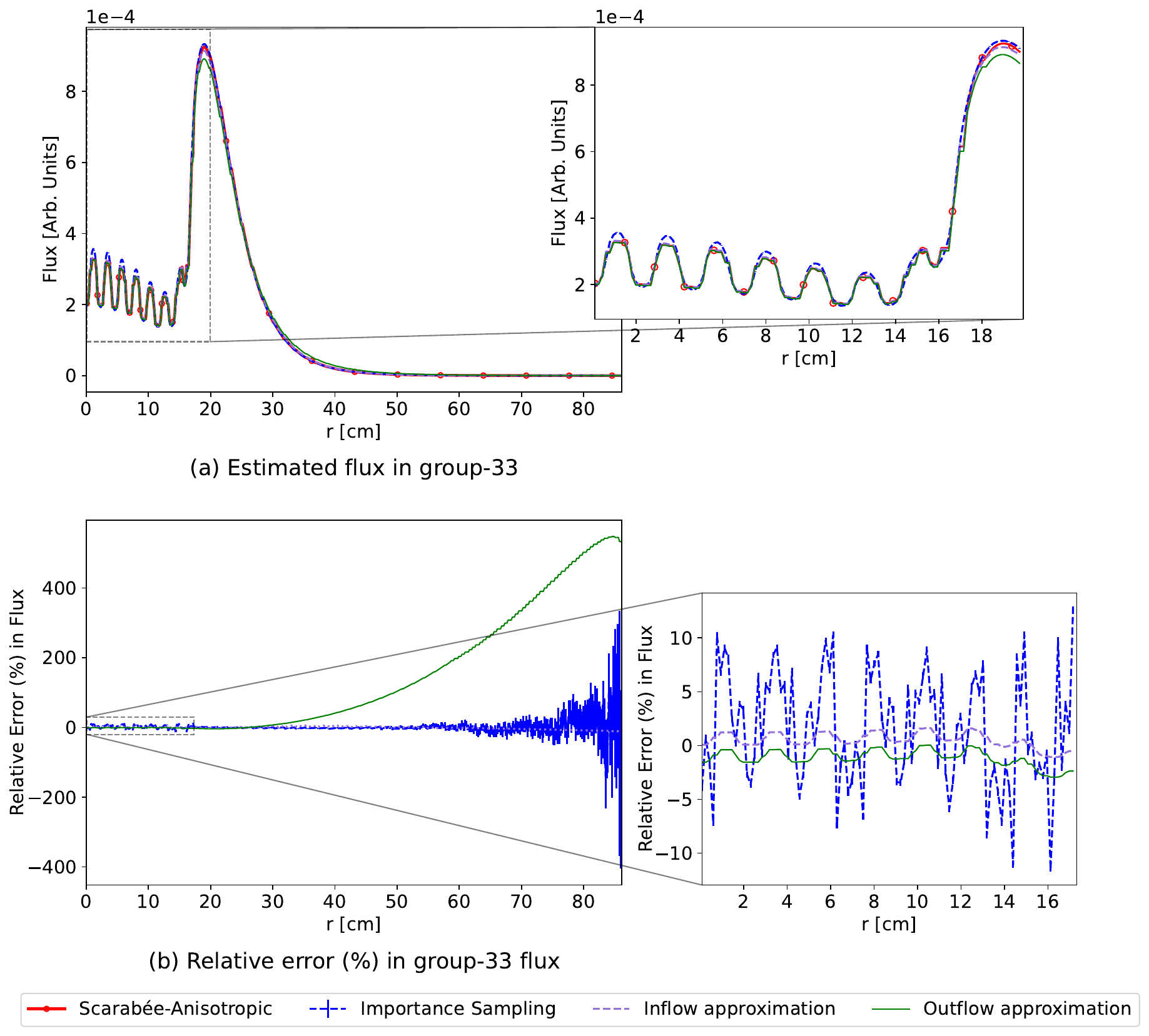}
    \caption{Group-33 estimated scalar flux and associated relative error (\%) along the diagonal of the modified RCF geometry by different methods }
    \label{fig:RCF-flux-comparision-group-33}
\end{figure}
\begin{figure}[!bt]
    \centering
    \includegraphics[width=1.\linewidth]{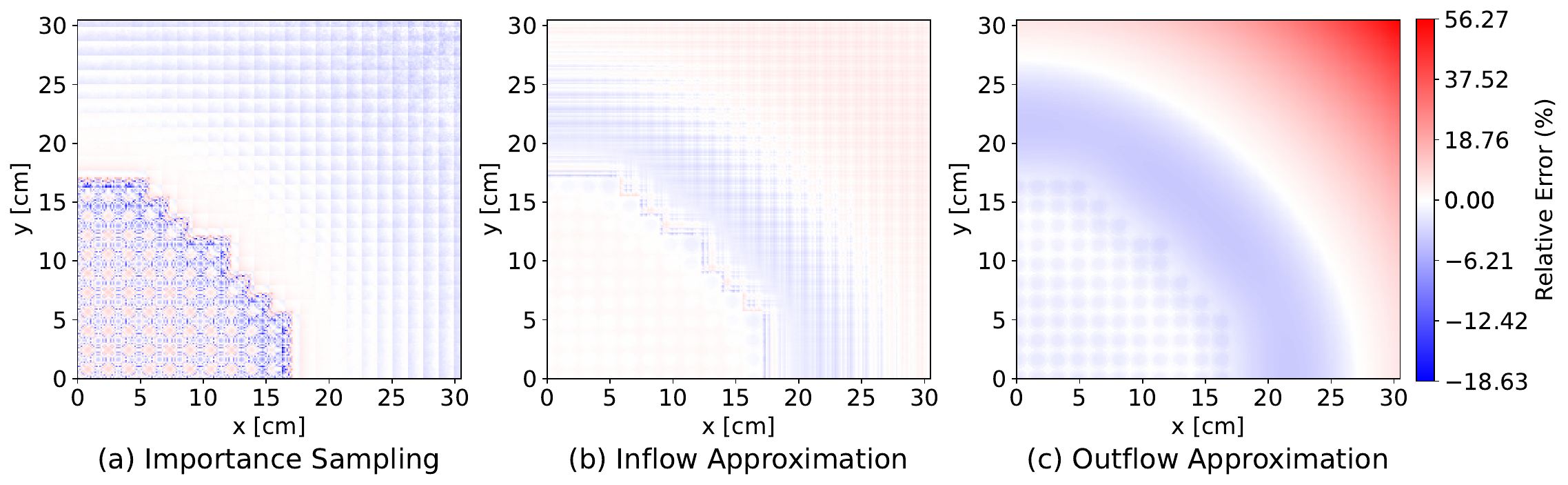}
    \caption{2D contour of relative error (\%) in scalar flux of group-33 calculated by different methods}
    \label{fig:RCF-flux-2D-contour-rel-error-group-33}
\end{figure}

Fig.~\ref{fig:RCF-flux-comparision-group-33}(a) and (b) show the scalar flux and relative errors, respectively, for group 33 along the diagonal of the core. With Fig.~\ref{fig:RCF-flux-comparision-group-33}(a), an enlarged portion is provided with the region over the fuel lattice. It can be seen in this case that all three MOC simulation results are in very good agreement with one another, but the Monte Carlo results from importance sampling give a higher estimate for the flux in the moderator between fuel pins. This is not really a bias in our importance sampling technique, but a result of the how the moderator region in pin cells was discretized into flat source regions. Like in the reflector region, this makes it difficult for the MOC simulations to accurately compute the flux between fuel pins. This finding is reinforced by Fig.~\ref{fig:RCF-flux-2D-contour-rel-error-group-33}(a) where the flux errors clearly correspond to the flat source regions, and there is a systematic overestimation in the moderator between fuel pins. It is therefore evident that Monte Carlo results with importance sampling are more accurate than the anisotropic MOC results. This fact is reassuring, however, as the Monte Carlo simulation should be more accurate than the deterministic solution, and we see that this is the case despite the presence of negative weights and the approximate weight cancellation. Far away from the center of the core, there is again a much higher variance in the Monte Carlo results, but this is again driven by few particles in the region and the presence of negatively weighted particles. Similar to the other examined energy groups, the bias in the outflow approximation increases far from the core center, and there is again an out-in bias.

\section{Conclusions}\label{sec:conclusions}
In the present study, the challenges associated with performing multigroup Monte Carlo simulations with anisotropic scattering has been considered. Scattering distributions are typically represented as low order Legendre series expansions, often only going up to the second or third order. With such few terms, it is often the case that the probability distributions for highly anisotropic energy transfers can have negative regions. This makes it impossible to apply standard Monte Carlo sampling methods to the distribution, as the cumulative distribution function is not monotonically increasing. Other Monte Carlo codes have traditionally either limited multigroup Monte Carlo capabilities to only isotropic scattering, or have employed a discrete angle approximation. In either case, this limits the physical accuracy of solution, and also makes it impossible to properly verify deterministic codes which can explicitly make use of Legendre expansions, regardless of wether or not the probability distributions are positive or negative.

In this study, we have suggested the used of importance sampling with weight cancellation to perform multigroup Monte Carlo simulations with anisotropic scattering on angular distributions with negative regions. With importance sampling, we can instead sample a scattering angle from an alternative, statistically valid, probability distribution, and then multiply the particle's statistical weight by a likelihood ratio. If the true scattering distribution for the sampled angle is negative, however, the particle's statistical weight can be come negative. This technique has been previously proposed by Brockmann \cite{brockmann1981}, but is never used in practice as the negative particle weights have historically been observed to prevent criticality simulations from converging. As part of this study, we were able to mathematically demonstrate why the negative weights introduced by importance sampling prevent convergence, and that weight cancellation algorithms as proposed by Belanger et al.\ can stabilize such simulations \cite{exact_weight_cancellation}.

The presented methodology of importance sampling with approximate regional weight cancellation was implemented in the Abeille Monte Carlo code \cite{abeille_repo} and verified against the two analytically known eigenvalue problems. The results from the Monte Carlo simulations are also compared against results obtained with the Scarabée lattice physics code \cite{scarabée}. The estimates of $\keff$ via importance sampling and approximate weight cancellation were found to be within one standard deviation of the known $\keff$ for both problems. The flux obtained by importance sampling was also in excellent agreement with results from Scarabée using anisotropic scattering treatments. Other approximated angular distributions which were purely positive and preserved the Legendre moments of the scattering distributions were also tested in Abeille. These approximate distributions were not able to reproduce the analytically known eigenvalues for the two problems, and the estimated flux had absolute relative errors of 4-6 \% in problem-37 which is a one-group infinite cylinder, and up to 0.8 \% in problem-71 which is a two-group infinite slab.

As a final demonstration, a realistic reactor problem was considered, examining a 2D model of RPI's Research Critical Facility (RCF). Multigroup cross sections were generated with continuous energy Monte Carlo simulations, going up to the third order Legendre moment scattering matrix. With this expansion, many different energy transfers had scattering distributions with negative regions. The multigroup Monte Carlo results obtained with Abeille were compared against deterministic results obtained with Scarabée using the method of characteristics based on spherical harmonics to treat anisotropic scattering. Overall, the estimation of $\keff$ and the scalar flux in all groups using Abeille was found to be in good agreement with the deterministic results. While some differences were visible, most could be attributed to the flat-source approximation used by Scarabée, which struggled to adequately capture the flux gradient in the moderator between fuel pins. There were also some tally bins which recorded an average negative flux in the reflector far from the core. This is due to the presence of negative weights of the simulation, and such instances of a negative flux only occurred in areas very difficult for particles to reach and had relative uncertainties greater than 100 \%.

Overall, these results indicate that importance sampling with weight cancellation is an excellent option for performing multigroup Monte Carlo simulations while maintaining anisotropic scattering. This is likely to be of use in verifying deterministic solvers, and could be beneficial for GPU based multigroup Monte Carlo implementations. One interesting observation made during this work was that that using a weight cancellation mesh where each fuel pin has its own cancellation bin seemed sufficient to eliminate observable biases in the scalar flux and eigenvalue. Future research will examine how refined the weight cancellation mesh must be to eliminate biases. The coarser the cancellation mesh, the more efficient the simulation will be. While a $5\times 5$ cancellation mesh was used for each pin cell in this study, the observation that a single mesh bin may be sufficient is promising for future work that relies on weight cancellation.

\section*{Acknowledgments}
This research used resources of the National Energy Research Scientific Computing Center (NERSC), a Department of Energy Office of Science User Facility using NERSC award ASCR-ERCAP0029095.

\end{document}